\documentstyle[epsf]{l-aa}

\begin{document}

\thesaurus{06(02.08.1, 08.02.1, 08.14.1, 13.07.1)}

\title{Coalescing neutron stars -- gravitational waves from
polytropic models} 

\author{M.~Ruffert\thanks{e-mail: {\tt mruffert@mpa-garching.mpg.de}}
\and M.~Rampp\thanks{e-mail: {\tt mrampp@mpa-garching.mpg.de}} 
\and H.-Th.~Janka\thanks{e-mail: {\tt thj@mpa-garching.mpg.de}} }
\institute{MPI f\"ur Astrophysik, Postfach 1523, 
D-85740 Garching, Germany}

\offprints{M.~Ruffert}


\maketitle

\begin{abstract}

The dynamics, time evolution of the mass dis\-tri\-bution, and
gravitational wave signature of coalescing neutron stars described by
polytropes are compared with three simulations 
published previously:  
(a) ``Run~2'' of Zhuge et al.~(1994), 
(b) ``Model~III'' of Shibata et al.~(1992), and
(c) ``Model~A64'' of Ruffert et al.~(1996).
We aim at studying the differences due to the use of different
numerical methods, different implementations of the gravitational wave
backreaction, and different equations of state.
We integrate the three-dimensional Newtonian equations of
hydrodynamics by the Riemann-solver based ``Piecewise Parabolic
Method'' on an equidistant Cartesian grid.

Comparison~(a) confronts the results of our grid-based PPM scheme with
those from an SPH code.
We find that due to the lower numerical viscosity of the PPM code,
the post-merging oscillations and pulsations can be followed for a
longer time and lead to larger secondary and tertiary maxima of the
gravitational wave luminosity and to a stronger peak of the
gravitational wave spectrum at a frequency of about $f\approx1.8$~KHz
when compared to the results of Zhuge et al.~(1994). 
In case~(b) two grid based codes with the same backreaction formalism
but differing hydrodynamic integrators and slightly different initial
conditions are compared.
Instead of rotationally deformed initial neutron stars we use
spherically shaped stars.
Satisfactory agreement of the amplitude of the gravitational
wave luminosity is established, although due to the different initial
conditions a small time delay develops in
the onset of the dynamical instability setting in when the two stars
come very close.
In~(c) we find that
using a polytropic equation of state instead of the high-density
equation of state of Lattimer \& Swesty~(1991)
employed by Ruffert et al.~(1996) does not change the overall
dynamical evolution of the merger and yields agreement of the
gravitational wave signature to within 20\% accuracy. 
Whereas the polytropic law describes the dynamical behaviour of the
bulk of the matter at and above nuclear density sufficiently well, we,
however, find clear differences of the structure and evolution of the
outer layers of the neutron stars where the stiffness of the equation
of state is largely overestimated.
This has important implications for questions like mass loss and disk
formation during the merging of binary neutron stars.


\end{abstract}

\section{Introduction}

Coalescing neutron stars are interesting and powerful sources
of gravitational waves.
These mergers --- together with black holes, rapidly rotating neutron
stars, and supernovae --- are among the most promising candidates
(Thorne 1992) to be detected with the large gravitational wave
experiments that will soon be operational or are currently planned,
LIGO (Abramovici et al.~1992), 
VIRGO  (Bradaschia et al.~1991), GEO600 (Danzmann et al.~1995a), 
and the space-based LISA (Danzmann et al.~1995b).
The detectability does not only depend on the strength of the
emission, but also on the frequency of the gravitational waves
relative to the resonance frequencies of the detectors
(Finn \& Chernoff 1993; Finn 1994). 
Only an a priori knowledge of the structure of typical gravitational
wave signals allows the extraction of faint signals from a very
noisy background.
Theoretically predicted wave templates have to be produced by
numerical simulations. 
Additionally to being sources for gravitational waves,
coalescing neutron stars (and its relative, the merger
of a neutron star with a black hole) remain the favorite cosmological
model for gamma-ray bursters (Dermer \& Weiler 1995)  
because the sources are known to exist, the right amount of energy is
involved, and the expected merger rates are in agreement with the
required gamma-ray burst rates.


The search for gravitational waves thus calls for models of coalescing
neutron stars that are numerically highly resolved and as realistic as
possible. 
Several groups have attempted such investigations and
have published results.
In a series of papers starting with Oohara \& Nakamura (1989) a
Japanese group used variations of a grid-based code to find the
gravitational wave luminosity for different conditions, e.g.~different
masses, initial separation or spins of the neutron stars (Shibata et
al.~1993, and references therein).
Zhuge et al.~(1994) concentrated on the energy spectrum that gives the
energy emitted in gravitational waves at different frequencies. 
They discussed the information about the dynamics of the coalescence
that may be extracted from a measured spectrum. 
Focussing on dynamics and stability, Rasio \& Shapiro (1994) found
that {\it triaxial} configurations are formed by the merging of
neutron stars with a sufficiently stiff polytropic equation of state. 
Non-axisymmetric structures occur for adiabatic exponents of 3 or
higher (polytropic index 0.5 or below).
In this case the peak amplitude of 
the gravitational wave emission is substantially larger and the 
emission proceeds for a longer time after the coalescence. 


Our treatment of the gravitational wave backreaction is identical to
what has been used by Shibata et al.~(1992).
We consider this procedure to be more accurate than the point-mass
quadrupole approximation used by Davies et al.~(1994) and 
Zhuge et al.~(1994). 
Their approximation was used when the neutron stars were still
separated and subsequently switched off when the merging took place.
Although Rasio \& Shapiro~(1994) investigated the stability of the
neutron star binary orbit in detail, they did not take into account
the gravitational radiation reaction at all.

The three investigations done by Davies et al.~(1994), 
Zhuge et al.~(1994) and Rasio \& Shapiro (1994) all used SPH codes,
whereas 
our simulations, as well as the ones by Shibata et al.~(1992), were
performed with explicit, Eulerian, finite-difference grid-based schemes.
The algorithm we use is based on the PPM-scheme of Colella \&
Woodward~(1984) which is a higher-order Godunov-type method in that it
locally solves Riemann problems. 
It is thus superior to a particle-based method like SPH in the
handling of shocks and also has a rather small numerical viscosity
compared to other Eulerian algorithms and SPH.

We chose to compare our numerical scheme with the following three
calculations. 
(1) ``Run~2'' of Zhuge et al.~(1994) was chosen since
in this model the neutron stars have masses and radii similar to those
in Ruffert et al.~(1996) (although the equations of state were different).
Additionally, the published detailed spectral analysis of the
gravitational wave emission is well suited for comparisons. 
(2) ``Model~III'' of Shibata et al.~(1992) was selected, because their
numerical scheme is grid-based and the algorithm to treat the
gravitational wave backreaction is similar to ours. 
Therefore differences of the results are associated with
differing hydrodynamical integrators and numerical resolution.
In both cases~(1) and~(2), polytopic equations of state were used which
we also applied in our comparative calculations.
(3) ``Model~A64'' of Ruffert et al.~(1996) served for comparison 
in order to investigate the differences caused by the use of a
polytropic equation of state instead of the more realistic equation of
state of Lattimer \& Swesty (1991) which allowed Ruffert et al.~(1996)
also to include the effects due to neutrino emission.

We did not select a model computed by Davies et al.~(1994) because
although their computations were done with a polytropic equation of
state, they did not give information about the gravitational wave
emission, since their formula was too crude.
Rasio \& Shapiro~(1994), on the other hand, were very careful in
constructing equilibrium models and in investigating the stability of
these models as 
well as the gravitational wave forms and luminosity.
Nevertheless, we preferred for our present investigations a comparison
with a run of Zhuge et al.~(1994), 
because (a) Rasio \& Shapiro~(1994) did not include any backreaction of
the gravitational waves, and (b) the spectra presented by 
Zhuge et al.~(1994) were better suited for comparisons.

Section~\ref{sec:numer} gives a brief description of the numerical
methods, introduces the set of computed models and details the 
initial conditions that our simulations are started from.
The results of our computations and the comparison with previous
published models are presented in the
three Sects.~\ref{sec:resultsZCM}, \ref{sec:resultsSNO}, and
\ref{sec:resultsRJS}, separately.
A summary follows in Sect.~\ref{sec:end}.

\begin{table*}
\caption[]{
Parameters of the three Models~ZCM, SNO, and~RJS.
$R$ is the initial radius of the two identical neutron stars,
$\rho_0$ their initial central density,
$a_0$ the initial center-to-center separation of the stars,
$M$ the mass of the neutron stars,
$K$ the polytropic constant,
$\Gamma$ the adiabatic exponent,
$L$ the grid size,
$N$ the number of zones per dimension,
$t_{\rm f}$ the total time interval of the run,
${\cal E}$ the total energy emitted in gravitational waves within 
time~$t_{\rm f}$,
$\widehat{\cal L}$ is the maximum gravitational-wave luminosity, and
$r\widehat{h}$ is the maximum amplitude of the gravitational waveforms
observed from a distance~$r$
}
\label{tab:models}
\begin{flushleft}
\tabcolsep=1.7mm
\begin{tabular}{lccccccccccccl}
\hline\\[-3mm]
Model & $R$ & $\rho_0$  & $a_0$& $M$ & $K$& $\Gamma$
      & $L$ & $N$ & $t_{\rm f}$
      & $\widehat{\cal L}$ & $r\widehat{h}$ & ${\cal E}$ & notes\\ 
      & km & $10^{14}\frac{{\rm g}}{{\rm cm}^3}$
      & km & $M_\odot$& & & km & & ms 
      & $10^{55}$erg & $10^4$cm & $10^{52}\frac{\rm erg}{\rm s}$
      & reference, model name\\[0.3ex] \hline\\[-3mm]
ZCM & 15 & ~6.4 & 45 & 1.4~ & $9.7\cdot10^4$ & 2.0~~
     & 88 & ~64 & 8.7 & 2.0 & 1.9 & ~0.88
     & Zhuge et al.~(1994), Run~2 \\
SNO & ~9 & 29.6 & 27 & 1.4~ & $3.5\cdot10^4$ & 2.0~~
     & 52 & 128 & 2.0 & 8.3 & 3.2 & 13.6~ 
     & Shibata et al.~(1992), Model~III \\
RJS & 15 & ~5.7 & 42 & 1.63 & 2.438 & 2.319
     & 82 & ~64 & 4.6 & 2.8 & 2.6 & ~2.07
     & Ruffert et al.~(1996), Model~A64 \\
\hline
\end{tabular}
\end{flushleft}
\end{table*}

\section{Computational procedure and initial conditions\label{sec:numer}}

The numerical procedures implemented to simulate our models of
coalescing neutron stars were described in detail 
in a first paper (Ruffert et al.~1996).
In the following, we shall specify only the differences of the treatment
of the presented polytropic models compared to what has been
described in Ruffert et al.~(1996).

\subsection{Hydrodynamics and equation of state}

The neutron stars are evolved hydrodynamically via the Piecewise
Parabolic Method (PPM) developed by Colella and Woodward (1984).
The code is basically Newtonian, but contains the terms necessary to
describe gravitational waves and their backreaction on the
hydrodynamical flow (Blanchet, Damour \& Sch\"afer 1990) via energy
and angular momentum loss.
The volume integral over the local emission and backreaction terms
yields the total gravitational wave luminosity.
All spatial derivatives necessary to compute the gravitational wave terms
are implemented as standard centered differences on the grid.

The Poisson equations in integral form, necessary for the calculation
of the gravitational potential as well as for the wave backreaction,
are interpreted as convolution
and calculated by fast Fourier transform routines:
non-periodic grid boundaries are enforced by zero-padding
(e.g.~Press et al.~1992; Eastwood \& Brownrigg 1979).
The accelerations that follow from the potential are added as source
terms into the PPM algorithm.
For models~ZCM and RJS the number of cubic grid zones in the orbital
plane is 64~by~64, while perpendicular to the orbital plane 16~zones
were used with the same zone size in all three dimensions.
We use double the amount of zones (128) in every spatial direction for
model~SNO in order to have a similar number of zones (121) as
Shibata et al.~(1992).
We take advantage of the symmetry about the orbital plane to actually
calculate only one side of the plane.

A polytropic equation of state,
\begin{equation}
 P = K \rho^\Gamma \quad,
\label{eq:Peos}
\end{equation}
is used which relates the density $\rho$ to the pressure $P$ via two
parameters, the polytropic constant $K$ and the adiabatic 
exponent $\Gamma= 1+\frac{1}{n}$, with $n$ being the polytropic index.
This equation with $K$ and $\Gamma$ chosen to be the same in the whole
star was employed to construct the initial model of the neutron star.
Since the PPM code is written in conservative form (not taking into
account source terms like gravity), the primary variables are the
density, the momentum, and the total energy.
One therefore has the choice of using either Eq.~\ref{eq:Peos} to
calculate the pressure from the density or the relation
\begin{equation}
 P = (\Gamma-1) e
\label{eq:Eeos}
\end{equation}
to deduce the pressure from the internal energy density $e$.
Analytically these formulations are equivalent, but numerical
errors may destroy this equivalence.
When shocks or viscous shear dissipation occurs in the hydrodynamic
calculations, the use of Eq.~\ref{eq:Eeos} is preferable because due
to the dissipative processes the parameter $K$ of Eq.~\ref{eq:Peos} is
not a strict constant in space and time during the simulations. 
If one postulates analytical equivalence of the pressure in
Eq.~\ref{eq:Eeos} with the polytropic equation of state,
Eq.~\ref{eq:Peos}, the generation 
of thermal energy in shocks manifests itself in a variation of $K$.
Eq.~\ref{eq:Peos} was used to construct the initial model and 
Eq.~\ref{eq:Eeos} to compute the hydrodynamic evolution.
Due to these simple equations of state,
effects of neutrino emission, absorption or annihilation were not
included in the present simulations.

\subsection{Initial setup\label{sec:initial}}

Our initial setup involves placing two identical spherical neutron
stars at some distance from each other and giving them an initial
velocity. 
We prescribe the initial orbital velocities of the coalescing neutron
stars according to the motions of point masses, as computed
from the quadrupole formula.
The tangential components of the velocities of the neutron star
centers correspond to Kepler velocities on circular orbits,
while the radial velocity components reflect the emission of
gravitational waves leading to the inspiral of the orbiting bodies
(for details, see Ruffert et al.~1996).
A spin of the neutron stars around their respective
centers is {\it not} added.
Thus the initial state of the neutron star binary is defined by the
following parameters: (a) center-to-center distance between the
neutron stars and (b) three out of the following five quantities:
mass, radius, and central density of the neutron star, and polytropic
constant $K$, and adiabatic exponent $\Gamma$ of the equation of state. 

\begin{figure*}
 \begin{tabular}{cc}
  \epsfxsize=8.8cm \epsfclipon \epsffile{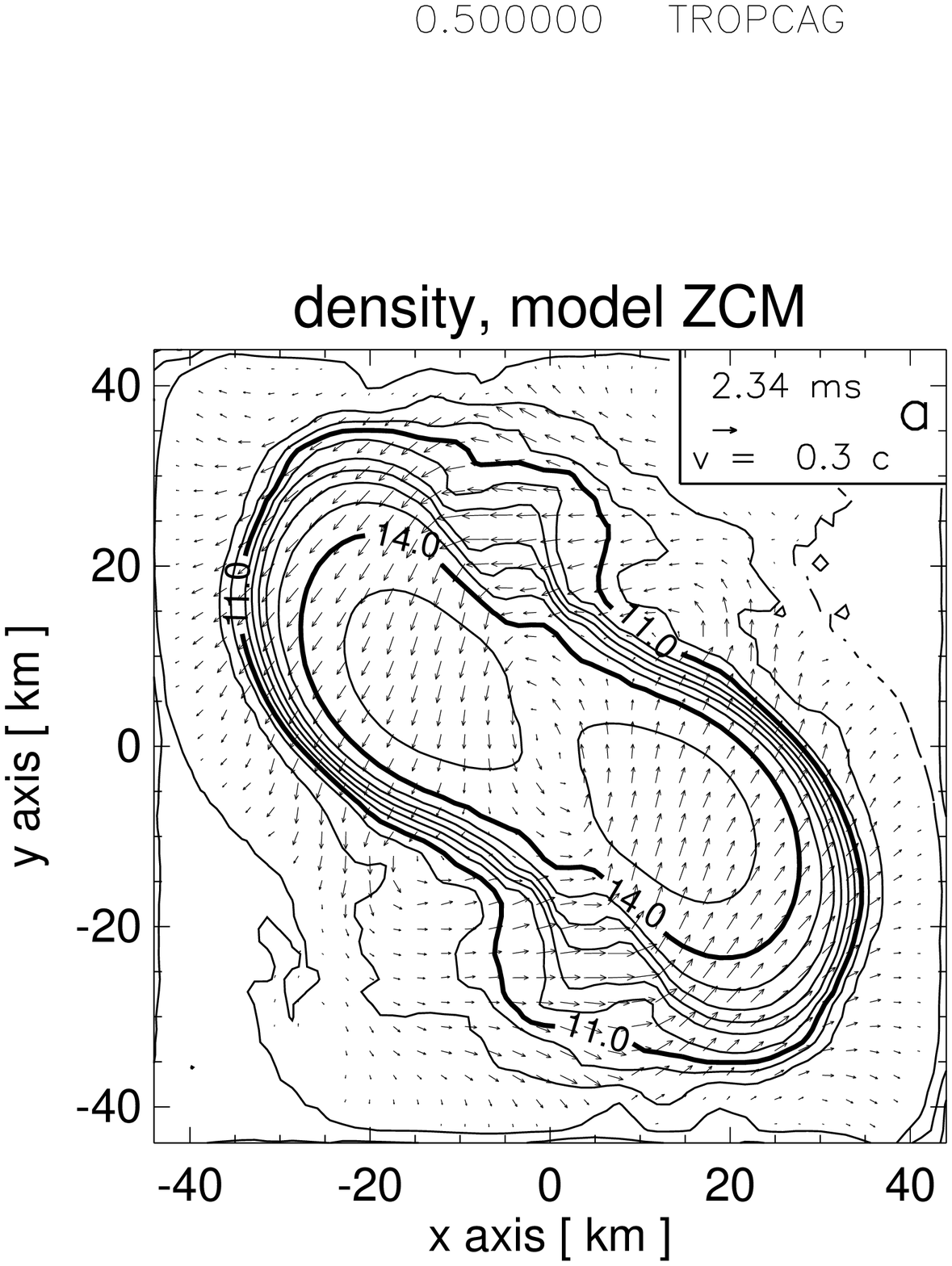} & 
  \epsfxsize=8.8cm \epsfclipon \epsffile{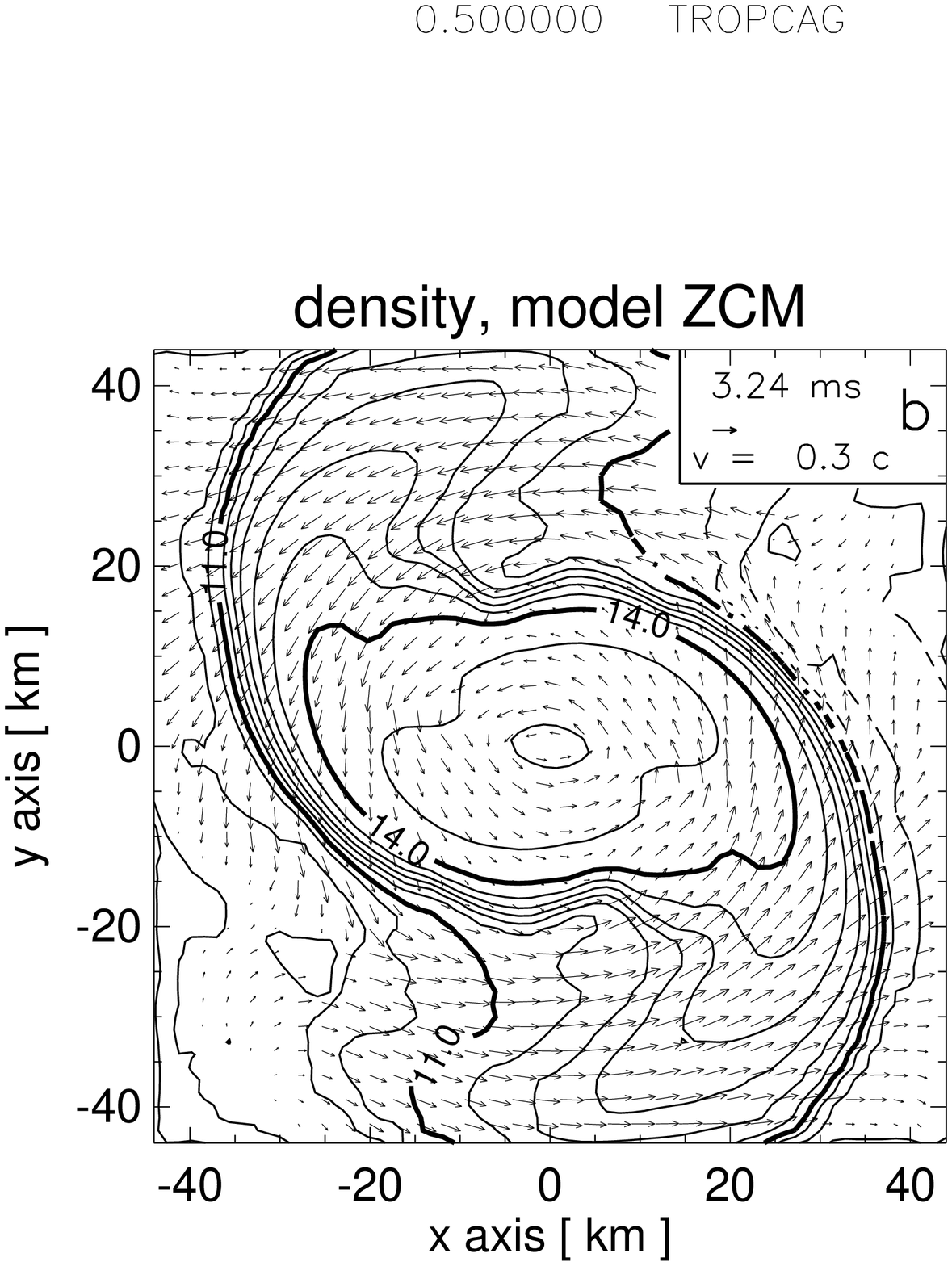} \\[-2ex]
  \epsfxsize=8.8cm \epsfclipon \epsffile{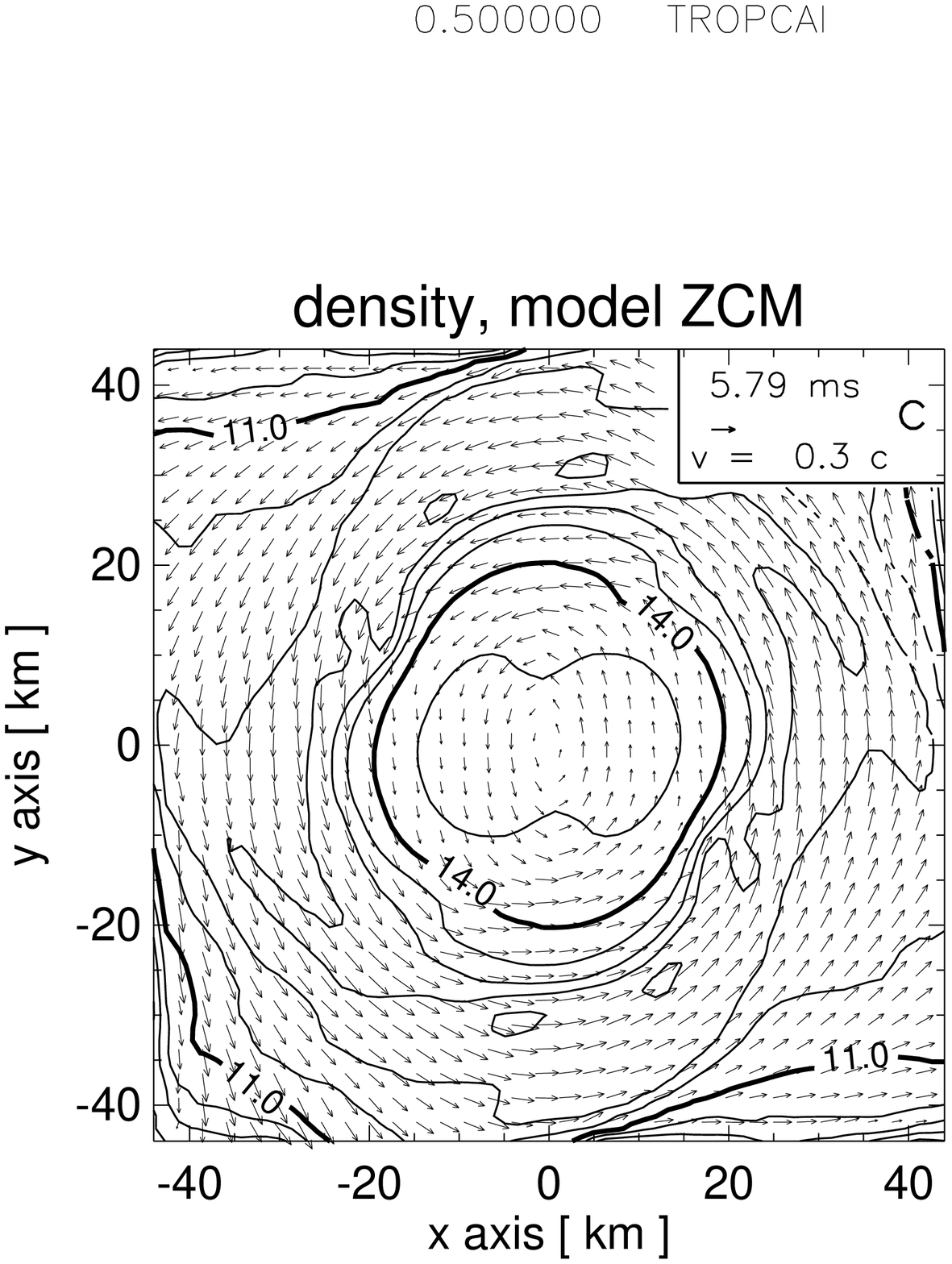} & 
  \epsfxsize=8.8cm \epsfclipon \epsffile{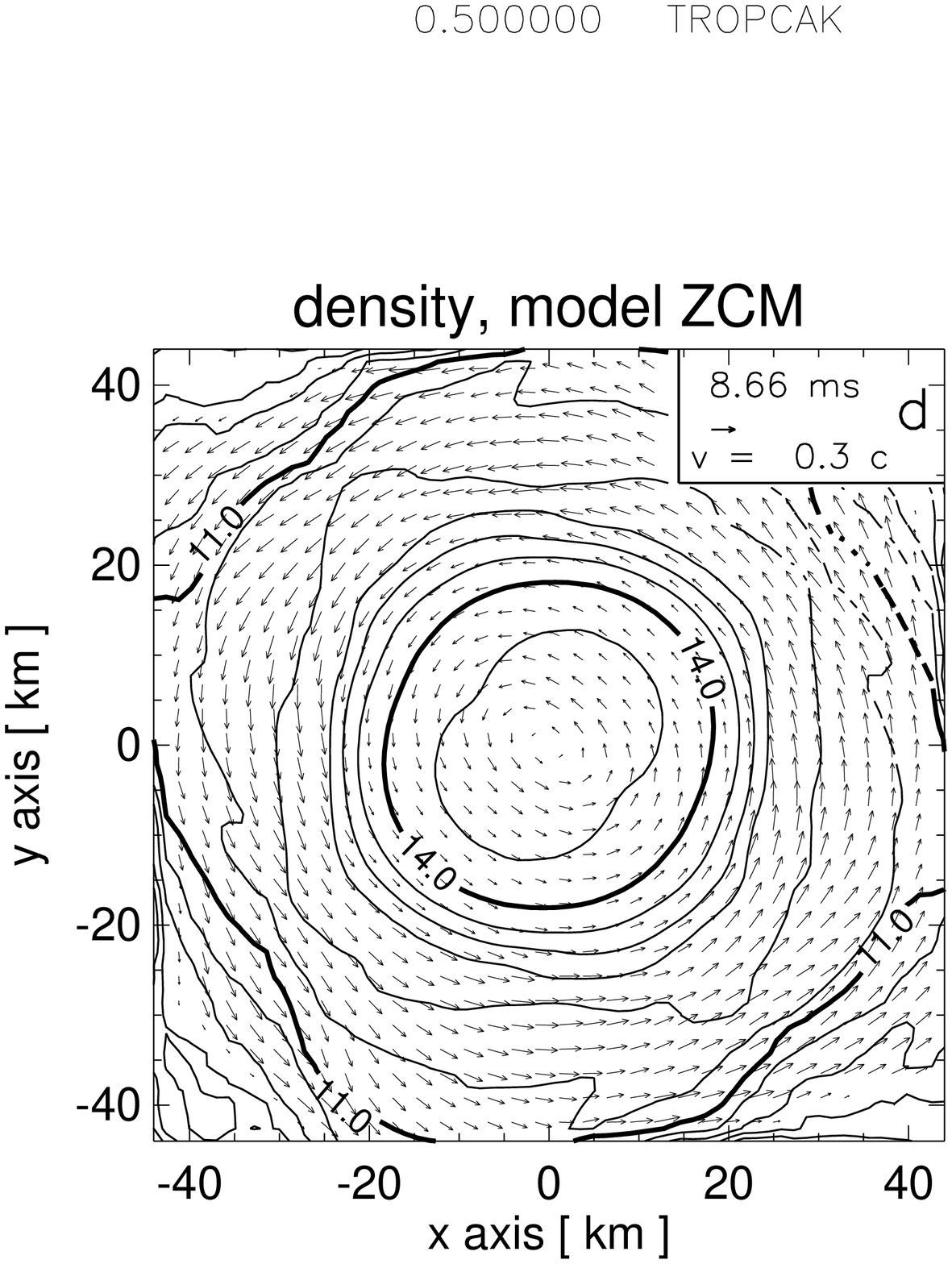} \\[-4ex]
 \end{tabular}
\caption[]{Cuts in the orbital plane of Model~ZCM at four instants in
time showing the density contours together with the velocity field.
The density contours are logarithmically spaced with intervals
of 0.5~dex. The density is measured in units of g/cm$^3$.
The bold contours are labeled with their respective values.
The legend at the top right corner of each panel gives the scale of
the velocity vectors and the time elapsed since the beginning of the
simulation
}
\label{fig:ZCMcont}
\end{figure*}

The initial distance at which the neutron stars are placed at the
beginning of our simulations is a compromise between physical
accuracy and computational resources.
On the one hand we would like to start with the neutron stars
at a large distance in order to correctly simulate the tidal
deformation they experience during inspiral.
However, the time to coalescence, which is proportional to the fourth
power of the distance, increases by a huge amount if the distance
between the neutron stars is only slightly increased.
On the other hand, tidal deformation studies (e.g.~Lai et al.~1994;
Reisenegger \& Goldreich 1994) show that tidal deformations for
extended objects are only around 20\% (for the principal axis) and
normal mode excitation of at most a few percent of the neutron star
radius at a distance of about 2.8 radii. 

For numerical reasons it is not possible to do the simulations with
arbitrary low density in the neutron star's environment.
So the density of the matter in the surroundings of the neutron stars is
set to $10^9$~g/cm$^3$.
To avoid this low-density matter being accelerated and falling onto the
neutron stars, the velocities and kinetic energy are reduced to zero
in zones in which the density is less than about $3\cdot10^9$~g/cm$^3$.
With a numerical grid of roughly 1~km spatial resolution one cannot
accurately represent the density decline near the surface of the
neutron stars.
We artificially soften the edge by imposing a maximal density change
of 2~orders of magnitude from zone to zone. 
The thickness of the surface layer thus results to about 3~grid zones.
To avoid short numerical time steps, we changed the constant $K$ of
the equation of state in the 3~surface zones such that 
the sound speed was constant within this broadened surface layer.
The same manipulation was applied for the surrounding gas at densities
of about $10^9$~g/cm$^3$.

We simulate three models with polytropic equations of state,
the parameters of which are listed in Table~\ref{tab:models}.
The initial conditions of each model are discussed in the respective
section of the corresponding model.
The calculations were performed on a Cray~J90~8/512, needed about
3.2~MWords of main memory and approximately 25~CPU-hours for
models ZCM and RJS, and 27~MWords and 140~CPU-hours for model~SNO.

\section{Results of Model ZCM\label{sec:resultsZCM}}

\subsection{Initial conditions}

We chose Run~2 of Zhuge et al.~(1994), because the mass and radius of
the neutron stars are closest to those of the stars which we have 
simulated in Ruffert et al.~(1996).
For Run~2, moreover, Zhuge et al.~(1994) show the gravitational
wave forms, the gravitational wave luminosity, and the energy spectrum.
This case yields a direct comparison between the SPH computations with
{\it ad hoc} gravitational backreaction done by Zhuge et al.~(1994)
and our Eulerian calculation of Model~ZCM where backreaction is included.

Model~ZCM has the same neutron star mass and radius as well
as an adiabatic exponent of $\Gamma=2$ as used by Zhuge et al.~(1994)
for their Run~2.
In order to save CPU time we place the neutron stars at an initial
distance of 45~km and {\it not} at the 60~km chosen by Zhuge et al.~(1994).
They were able to follow the inspiral from such a large distance
because their SPH simulation was done with a comparatively small
number of particles ($N=1024$) which is computationally much cheaper
than our grid-based simulations.
Unfortunately, they do not show a plot of the neutron star separation
as a function of time for Run~2.
Without this relation the connection between the temporal evolution of
their model and ours is not determined precisely.
Thus we link our time to theirs by matching both calculations at the
position of the first maximum of the gravitational wave luminosity
(see Fig.~\ref{fig:ZCMgrlum}) independent of how long it takes to
spiral in from the initial separation of 60~km (used by 
Zhuge et al.~1994) to our initial separation of~45~km.

However, the information given by Zhuge et al.~(1994)
for their Run~1 can be used to interpret their Run~2.
From their Fig.~10 one can infer at which time the separation
of their neutron stars is $3R$, and then inspect their Fig.~1a to get
an impression of the tidal deformations at that time:
tidal bulges are visible but not dominant, in agreement with the
analytical estimate of 20\% deformation of the principal axis at a
distance of $2.8R$ (see Sect.~\ref{sec:initial}).
We did not include these bulges in our initial setup of the neutron
stars.

\begin{figure}
 \epsfxsize=8.8cm \epsfclipon \epsffile{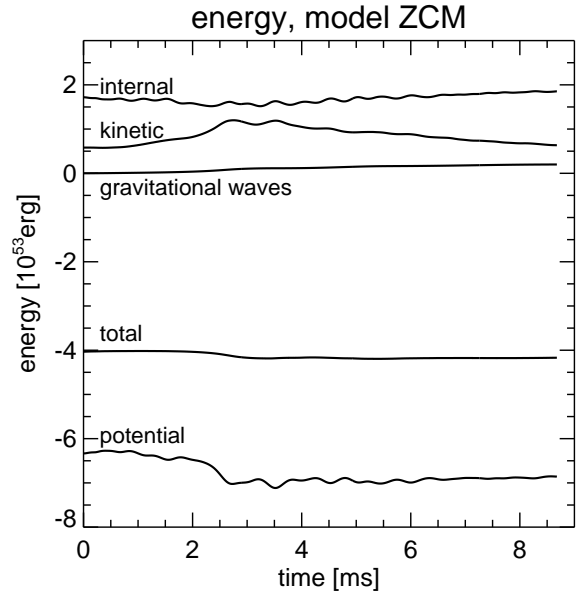}
\caption[]{Kinetic energy, internal energy, gravitational potential
energy, and emitted gravitational wave energy as functions of time for
Model~ZCM. The total energy contains all individual energies 
}
\label{fig:ZCMenergy}
\end{figure}

\begin{figure}
 \epsfxsize=8.8cm \epsfclipon \epsffile{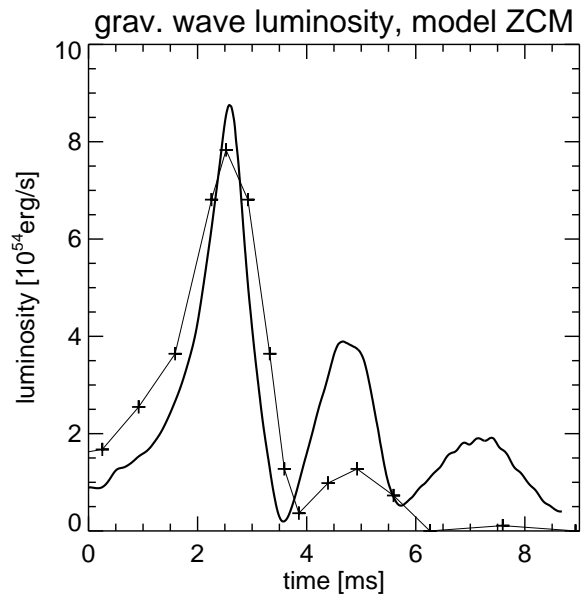}
\caption[]{The gravitational wave luminosity
as a function of time for Model~ZCM (solid line).
The crosses represent the values of the gravitational wave luminosity
taken from Fig.~16a in Zhuge et al.~(1994) for their Run~2; they are
connected by straight lines
}
\label{fig:ZCMgrlum}
\end{figure}

\begin{figure}
\epsfxsize=8.8cm \epsfclipon \epsffile{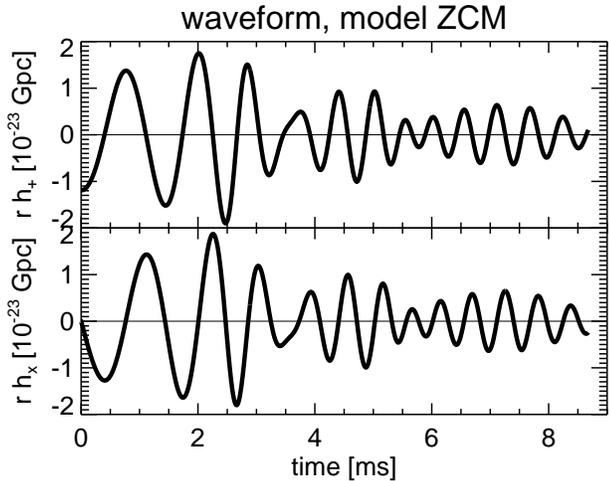}
\caption[]{The gravitational wave forms, $h_+$ and $h_\times$,
for Model~ZCM}
\label{fig:ZCMform}
\end{figure}

\subsection{Dynamical evolution}

Snapshots of the density distribution in the orbital plane together
with the velocity field are shown for four times in 
Fig.~\ref{fig:ZCMcont}.
These plots give an impression of how the merging process proceeds.
Within less than one orbital period the merging is well in progress
(Fig.~\ref{fig:ZCMcont}a).
The typical transient spiral-arm pattern develops
(Fig.~\ref{fig:ZCMcont}b) due to tidal and centrifugal forces.
One notices that for a fairly long time (for over 2~ms after $t=4$~ms;
e.g.~Fig.~\ref{fig:ZCMcont}c) the merged object has sort of a ``double 
core'' structure, visible by the dumb-bell shaped central contours.
The quadrupolar deformation  seems to be less pronounced in the
model of Zhuge et al.~(1994).
It is damped only slowly and indications of it are still present when
we stop our simulation at $t=8.7$~ms (Fig.~\ref{fig:ZCMcont}d).

The temporal evolution of different energies is shown in
Fig.~\ref{fig:ZCMenergy}.
One notices that the coalescence is very smooth and that the potential
energy decreases only by about 15\%.
The internal energy first decreases
a bit due to tidal stretching, then increases again because some
kinetic energy is dissipated into thermal energy,
indicating the presence of shocks and friction.
During the spiral-in and merging the maximum density decreases 
monotonically from roughly $6\cdot10^{14}$~g/cm$^3$ to
$5\cdot10^{14}$~g/cm$^3$, because of tidal stretching before
coalescence and internal ``heating'' due to shear effects which
dissipate kinetic energy during coalescence.\footnote{When the
hydrodynamics code uses the relation of Eq.~\ref{eq:Eeos}, heating
means a change of the constant $K$ in the polytropic equation of state,
Eq.~\ref{eq:Peos}.}
The total energy is conserved to 2\% compared to the maximum potential
energy.

\subsection{Gravitational wave forms and luminosity}

The temporal structure of the gravitational wave luminosity
is shown in Fig.~\ref{fig:ZCMgrlum} (the time integral of which can be
seen in Fig.~\ref{fig:ZCMenergy}).
The luminosity during the main merging event is of similar size in our
Model~ZCM and in Run~2 of Zhuge et al.~(1994).
Nevertheless, significant differences are visible:
the maximum at $t\approx2.5$~ms in our model is 10\% higher and the
width of the main peak about 40\% narrower.
Note that due to the difficulty of comparing the times for both
models, the temporal positions of the first luminosity maxima were
chosen to be the same and a numerical time shift of the maxima thus 
cannot be revealed. 
The largest difference of the gravitational wave luminosity appears at
the secondary maximum at $t\approx4.6$~ms after the actual merging
event. 
In our simulations this maximum is a factor
of 2.5 higher than  that shown in Zhuge et al.~(1994).
In addition, our Model~ZCM displays a tertiary maximum at
$t\approx7.3$~ms, with the time intervals between the maxima being
roughly 2.5~ms. 
The much more prominent secondary and tertiary maxima in our Model~ZCM
are due to the fact that the rotating merged object has a strongly
quadrupolar structure and performs oscillations and pulsations:  
two revolving ``subcores'' can be seen in Fig.~\ref{fig:ZCMcont}
which continue to emit gravitational
waves for as long as they are present and rotate.

The integrated energy emitted in gravitational waves is
$2.0\cdot10^{52}$~erg which corresponds to 0.79\% of $Mc^2$, $M$ being
the mass of one neutron star.
If one extracts the analogous values for the gravitational wave energy
from Fig.~16b in Zhuge et al.~(1994) for the time interval shown in
Fig.~\ref{fig:ZCMgrlum}, one obtains 0.8\% of $Mc^2$, in very good
agreement with our Model~ZCM, because the smaller peaks of the
gravitational wave luminosity in the model of Zhuge et al.~(1994) are
compensated by their larger widths.

For completeness we also show in Fig.~\ref{fig:ZCMform} the
gravitational wave forms for Model~ZCM, which can 
be compared with Fig.~15 of Zhuge et al.~(1994).
After the maximum amplitude is reached, the wave form of model~ZCM is
damped by roughly a factor of two within~10 further periods, while
the damping factor is about~20 in the simulations of 
Zhuge et al.~(1994). 
This difference reflects the motion of the two persisting ``subcores''
in our model.

\begin{figure*}
 \begin{tabular}{cc}
 \put(2.5,1.6){{\large $t=2.34$ms}}
  \epsfxsize=8.8cm \epsfclipon \epsffile{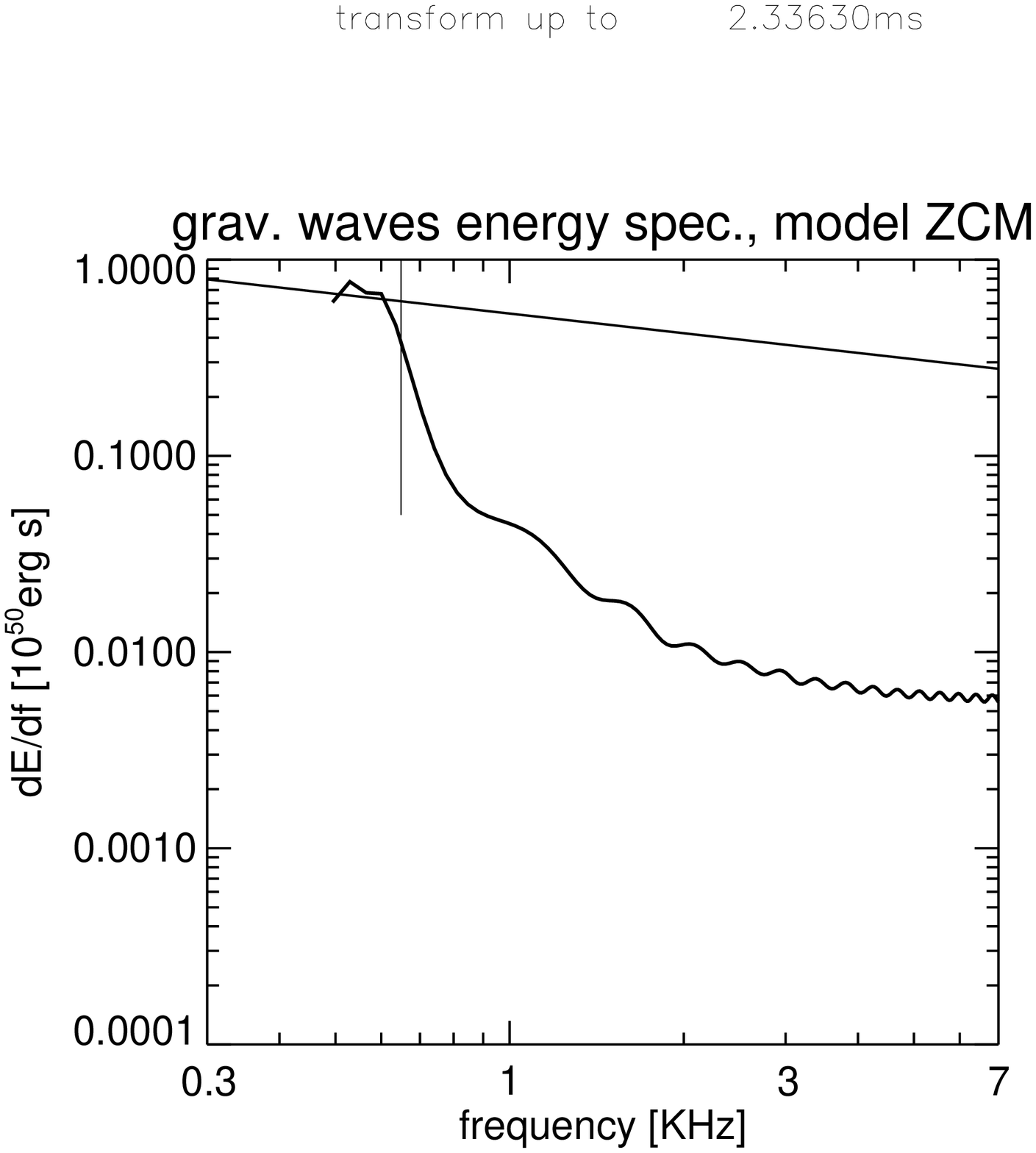} & 
 \put(2.5,1.6){{\large $t=3.24$ms}}
  \epsfxsize=8.8cm \epsfclipon \epsffile{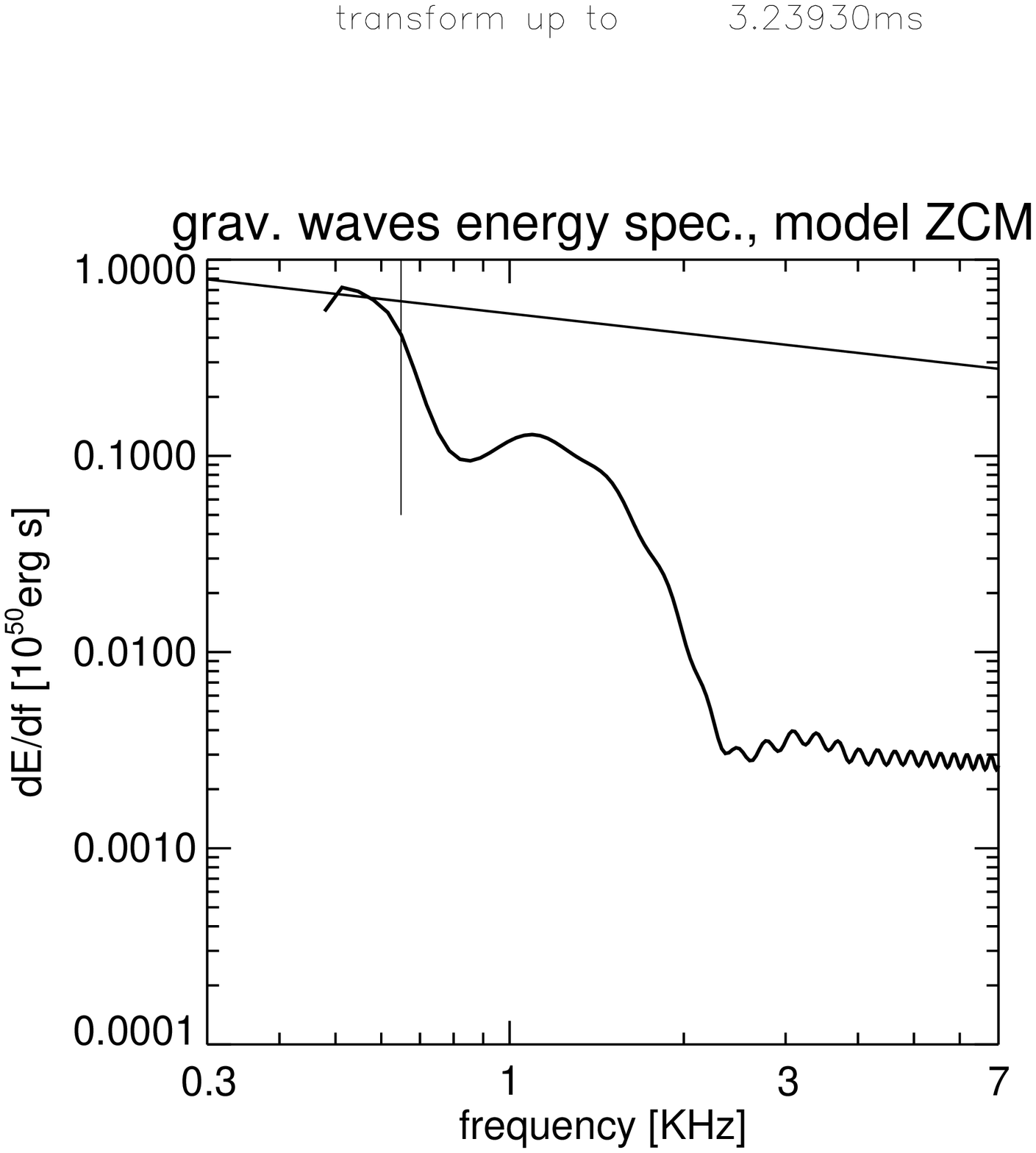} 
      \\[-2ex]
 \put(2.5,1.6){{\large $t=5.79$ms}}
  \epsfxsize=8.8cm \epsfclipon \epsffile{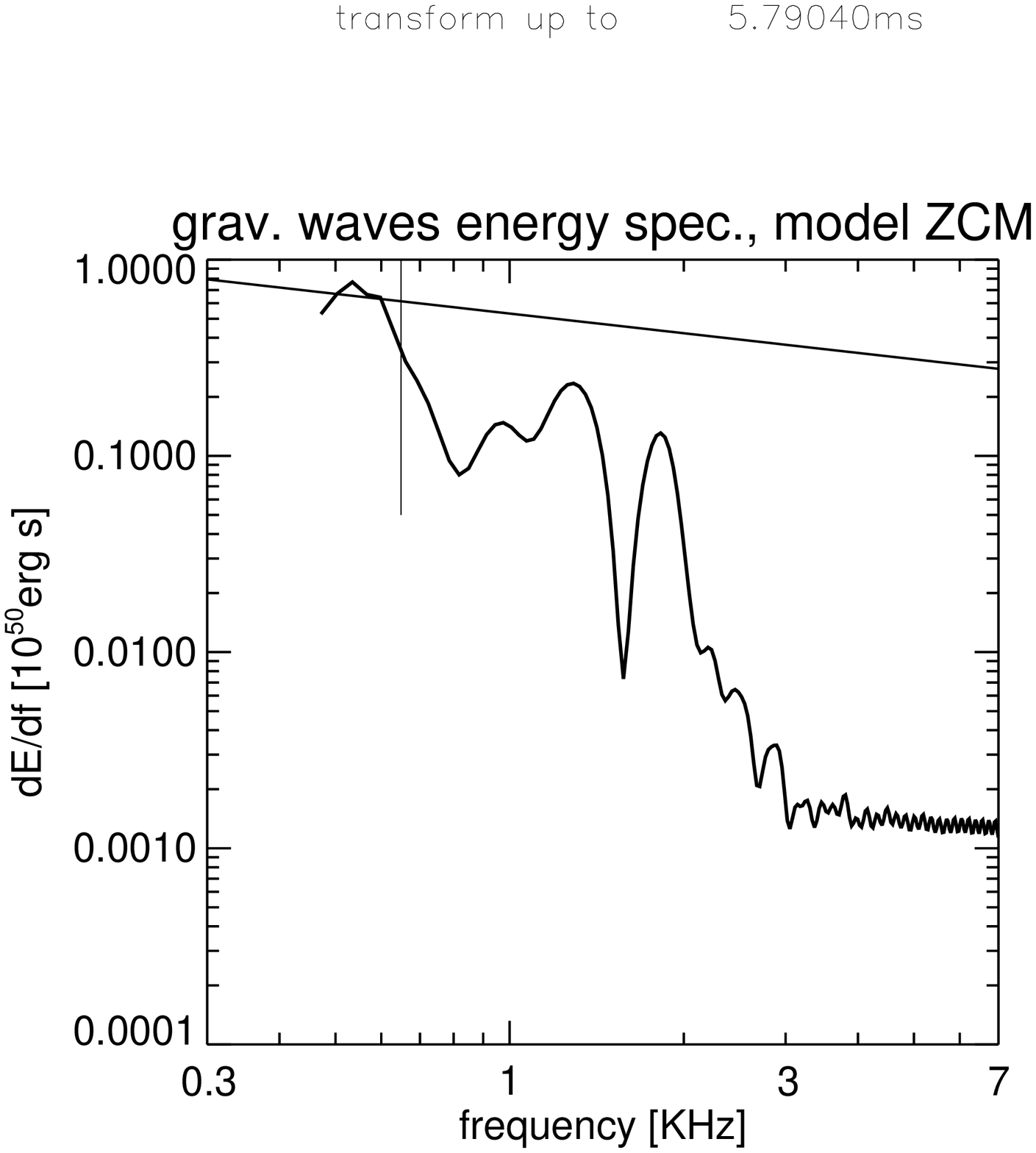} & 
 \put(2.5,1.6){{\large $t=8.66$ms}}
  \epsfxsize=8.8cm \epsfclipon \epsffile{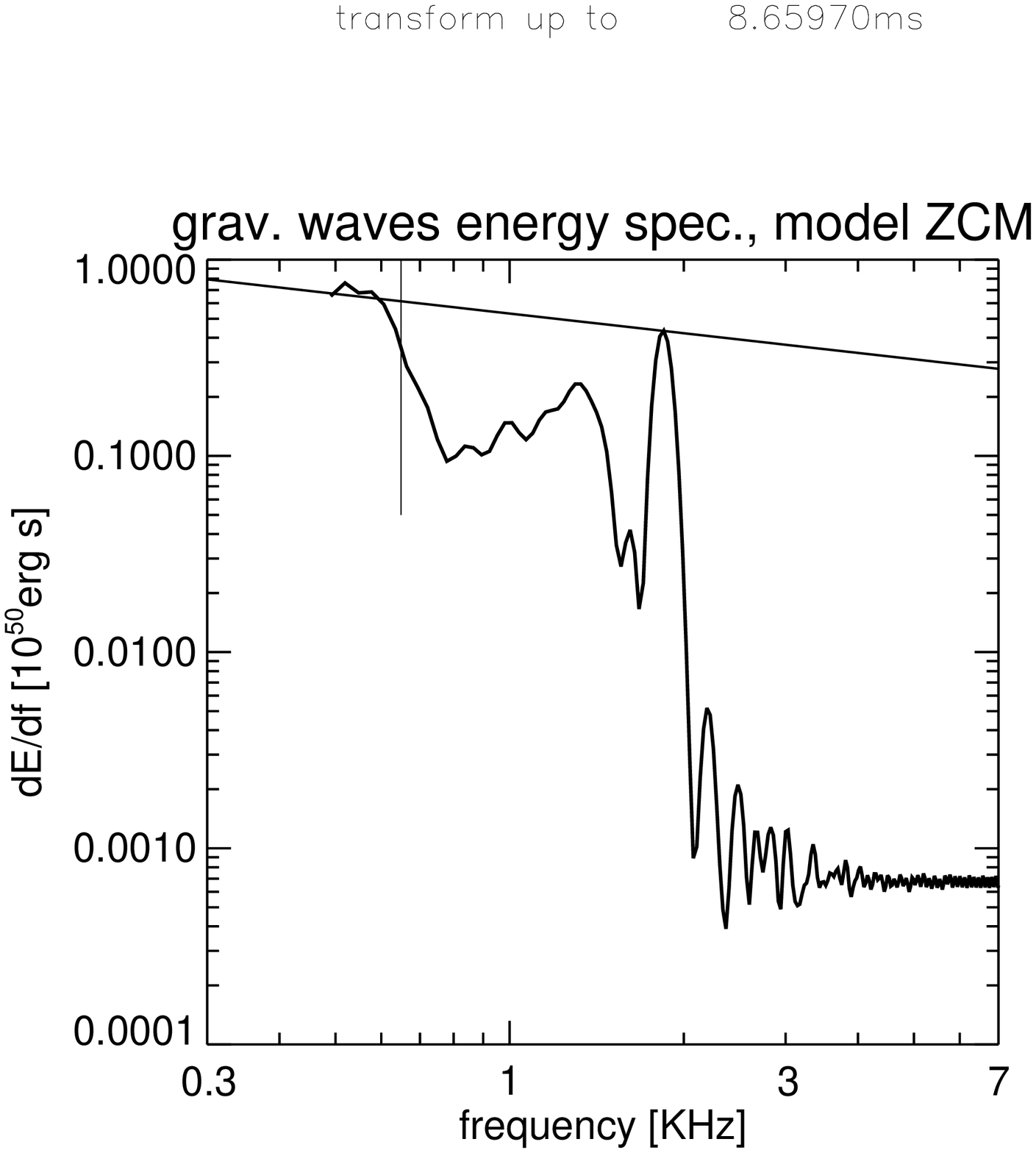} 
     \\[-4ex]
 \end{tabular}
\caption[]{Energy spectrum of gravitational waves emitted in Model~ZCM.
The times until which the Fourier transforms are performed are
indicated in  the lower left corner of the panels.
They correspond to the snapshots of the density distributions shown in
Fig.~\ref{fig:ZCMcont}.
The straight, downward sloping line is the spectrum of a point-mass
binary.
The vertical lines indicate the frequency corresponding to the orbital
frequency of two point masses at the initial distance of the neutron
stars in our numerical model
}
\label{fig:ZCMspec}
\end{figure*}

\subsection{Gravitational wave spectrum}

The cumulative emission of gravitational wave energy as a function of 
frequency is shown in Fig.~\ref{fig:ZCMspec}.
In the upper part of each panel, the downward sloping straight line
represents the energy loss per unit frequency interval of a point-mass 
binary.
The frequencies to the left of the vertical line correspond to the
wave frequencies that are emitted before the start of the numerical 
modelling, i.e.~for times $t<0$.
We produce a combined wave by using the quadrupole moments for a
point-mass binary for $t<0$ and taking the numerically
obtained quadrupole moments at $t>0$ (in analogy to Zhuge et al.~1994).
The combined wave is then Fourier analysed up to the times given 
in the lower left corners of the panels of Fig.~\ref{fig:ZCMspec}.

At low frequencies the
energy spectrum calculated for the combined wave fits very well
to what is expected from the point-mass approximation.
Although the merged object in the simulation radiates
gravitational waves for a longer time than the point-masses, the
very much smaller amplitudes associated with the coalescence of
extended neutron stars result in less energy emitted at almost all
higher frequencies. 
By Fourier transforming the signal of the combined wave until
different times $t>0$, 
we are able to roughly locate the moments when the peaks are produced. 

At around $t\approx2.3$~ms the spectrum is structureless.
A peak at roughly $f\approx1.14$~KHz starts to form at
$t\approx3.24$~ms and has saturated by $t\approx5.79$~ms and shifted
to around $f\approx1.22$~KHz. 
A second peak around $f\approx1.67$~KHz is visible at this time but
continues to grow until $t\approx8.7$~ms and to shift to
$f\approx1.79$~KHz. 
The timing of the two peaks corresponds very well to the three maxima
of the gravitational wave luminosity
(Fig.~\ref{fig:ZCMgrlum}): the peak at $f\approx1.22$~KHz is formed
by the first outburst of gravitational wave emission and the 
$f\approx1.79$~KHz peak by the
second and third phases of strong emission.
Fig.~17 in Zhuge et al.~(1994) displays the corresponding wave spectrum
for their model.
One notices that (a) we confirm their dip below 1~KHz, (b) we also see
the slight rise above 1~KHz, (c) they quote a value for a peak 
at 1.75~KHz which we indeed find at 1.79~KHz, and (d) their 1.75~KHz
peak is significantly smaller than the one of our model.
Thus the overall shape of the spectrum of our Model~ZCM is similar 
to the spectrum of Run~2 of Zhuge et al.~(1994).
The visible differences, in particular the high frequency spectrum and
the spectral maximum at $f\approx1.75$~KHz (point~(d) above), 
are associated
with the fact that due to the quadrupolar deformation of the merged
object in our simulation the emission of gravitational waves continues
for a longer time than in Run~2 of Zhuge et al.~(1994).
This emission of waves at higher frequencies is an indication that
the merged object becomes more compact with time. 
The gravitational waves emitted between 4~ms and 8~ms
(Fig.~\ref{fig:ZCMgrlum}) have a frequency of roughly 1.8~KHz,
while those originating between 2~ms and 4~ms have around~1.2~KHz.

\begin{figure*}
 \begin{tabular}{cc}
  \epsfxsize=8.8cm \epsfclipon \epsffile{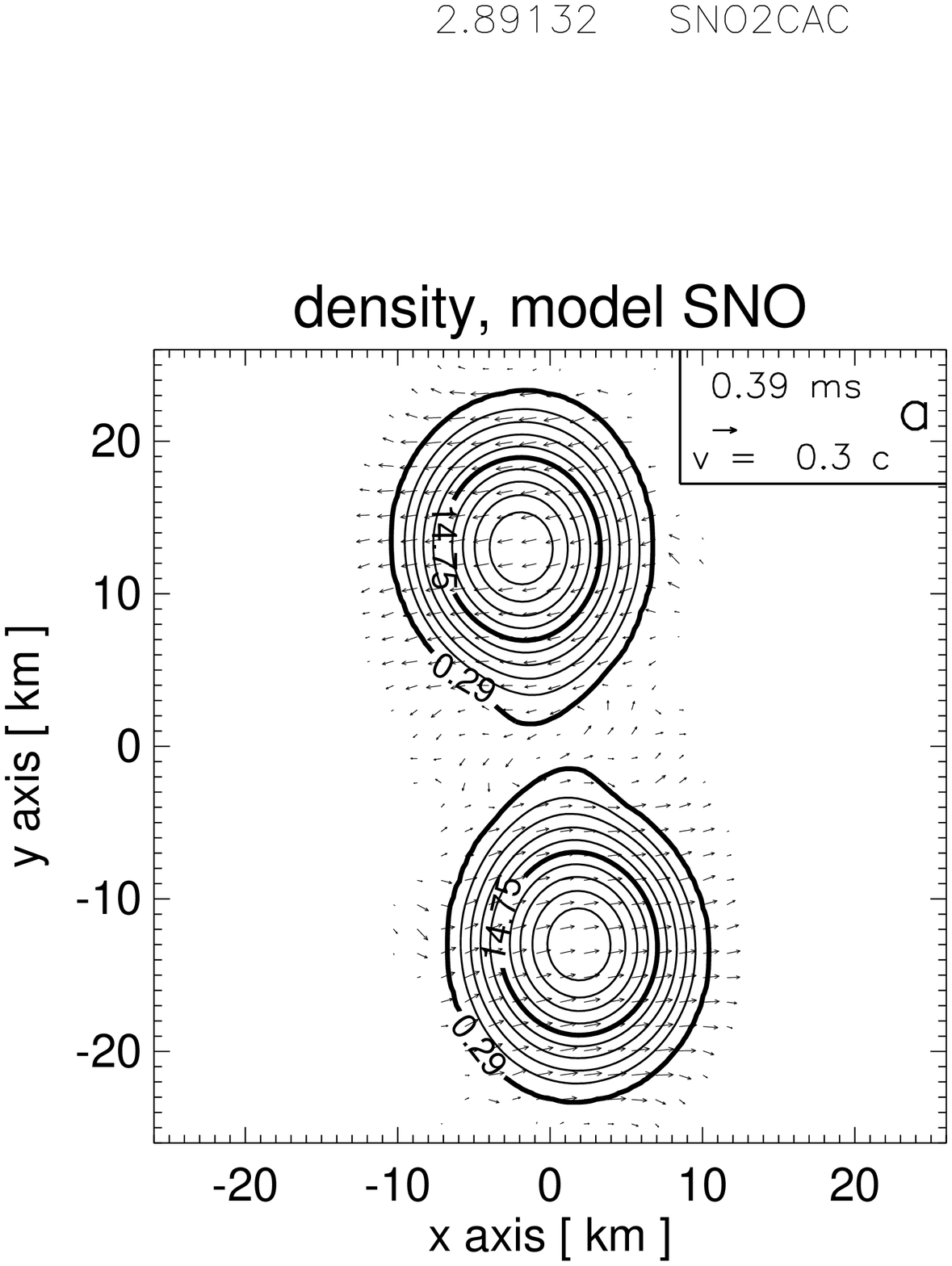} & 
  \epsfxsize=8.8cm \epsfclipon \epsffile{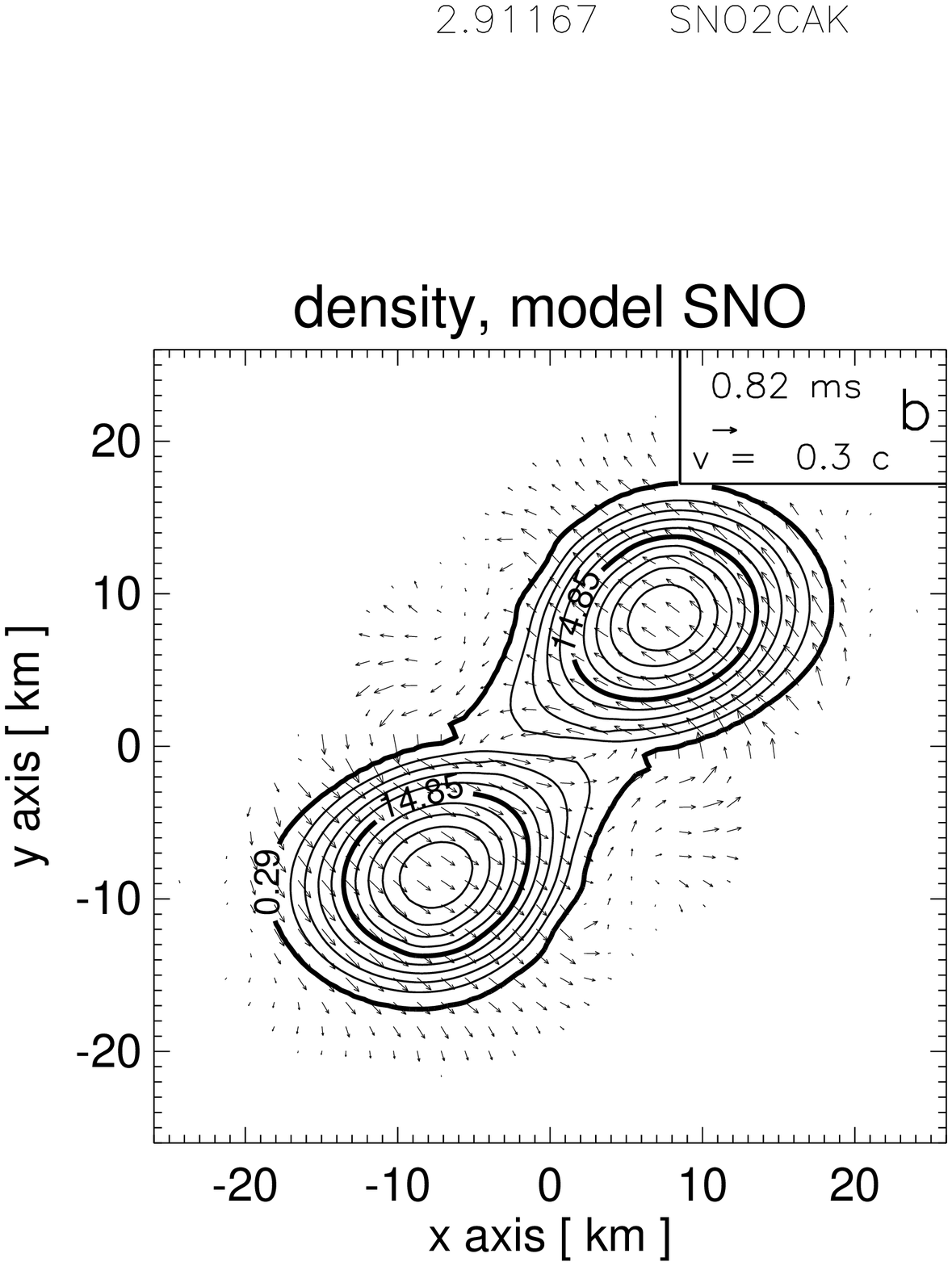} \\[-2ex]
  \epsfxsize=8.8cm \epsfclipon \epsffile{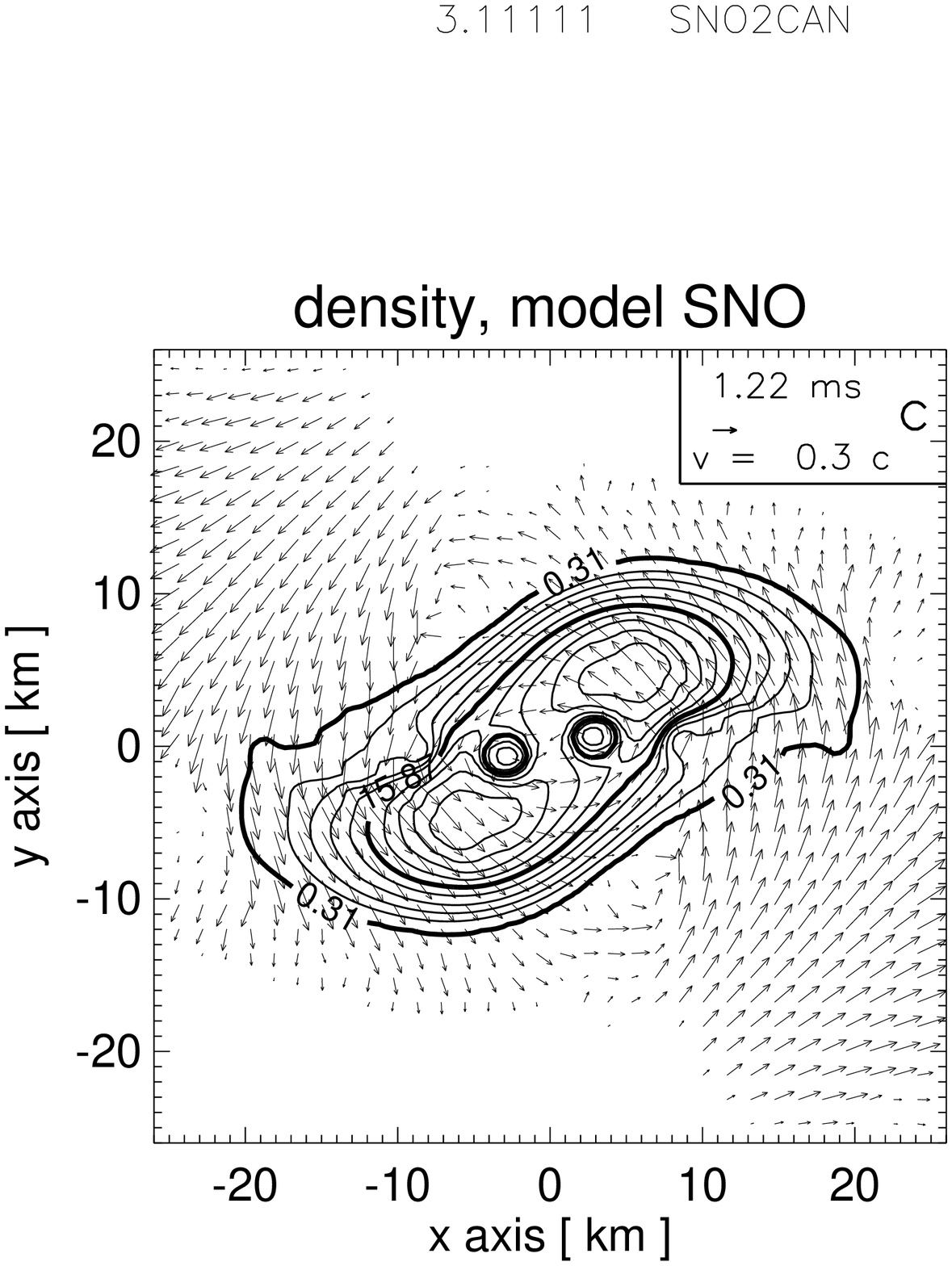} & 
  \epsfxsize=8.8cm \epsfclipon \epsffile{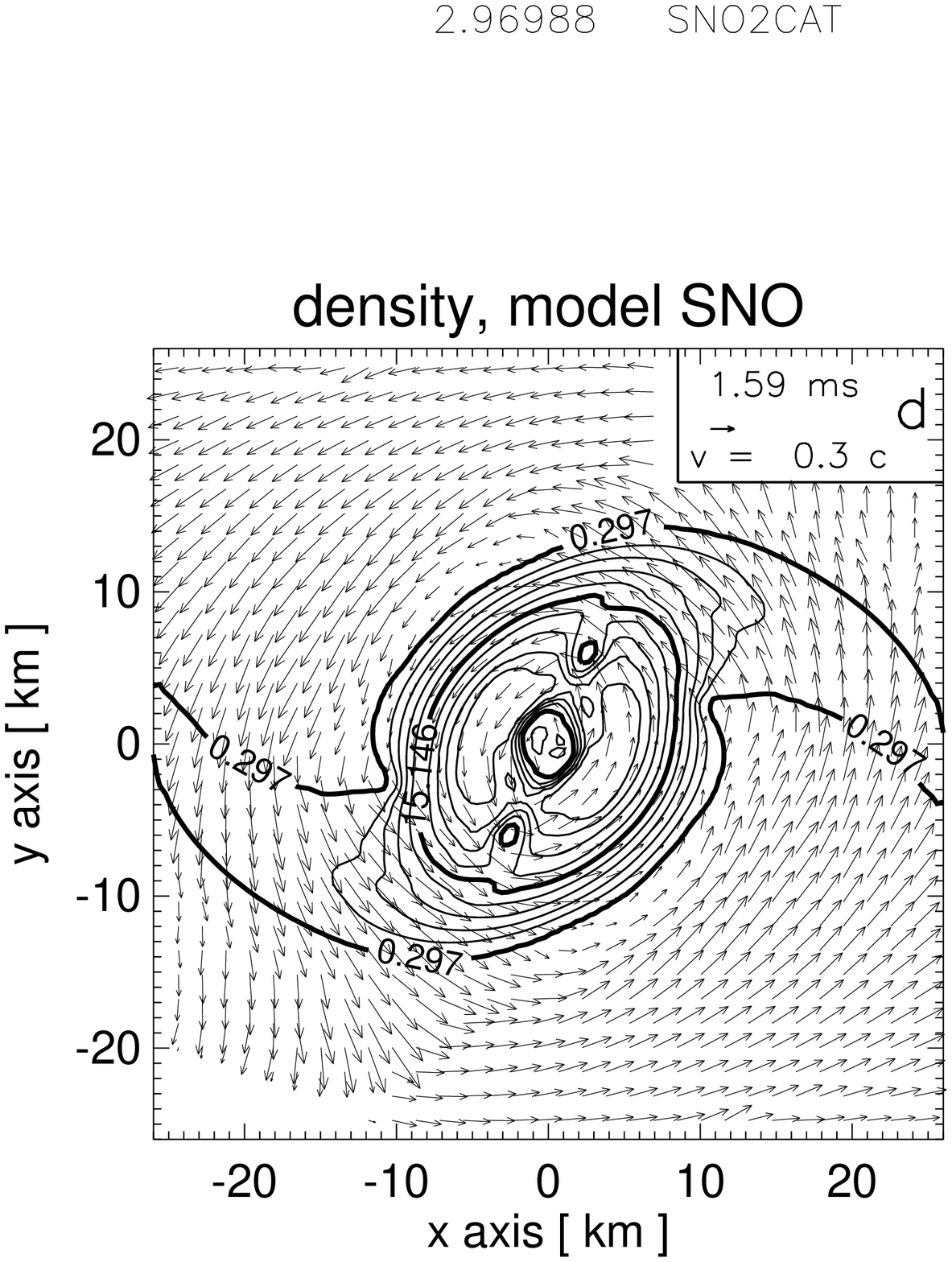} \\[-2ex]
  \epsfxsize=8.8cm \epsfclipon \epsffile{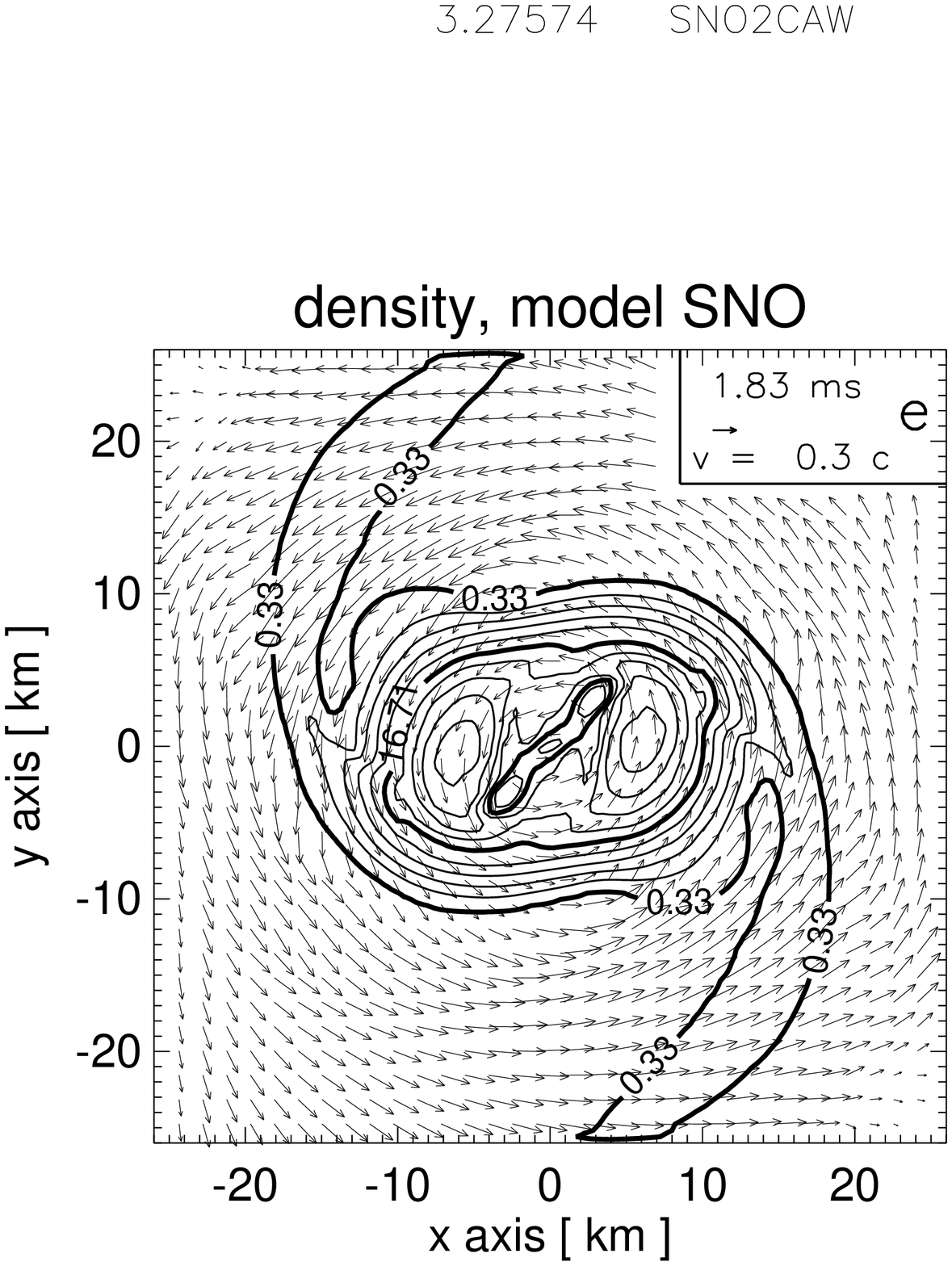} & 
  \epsfxsize=8.8cm \epsfclipon \epsffile{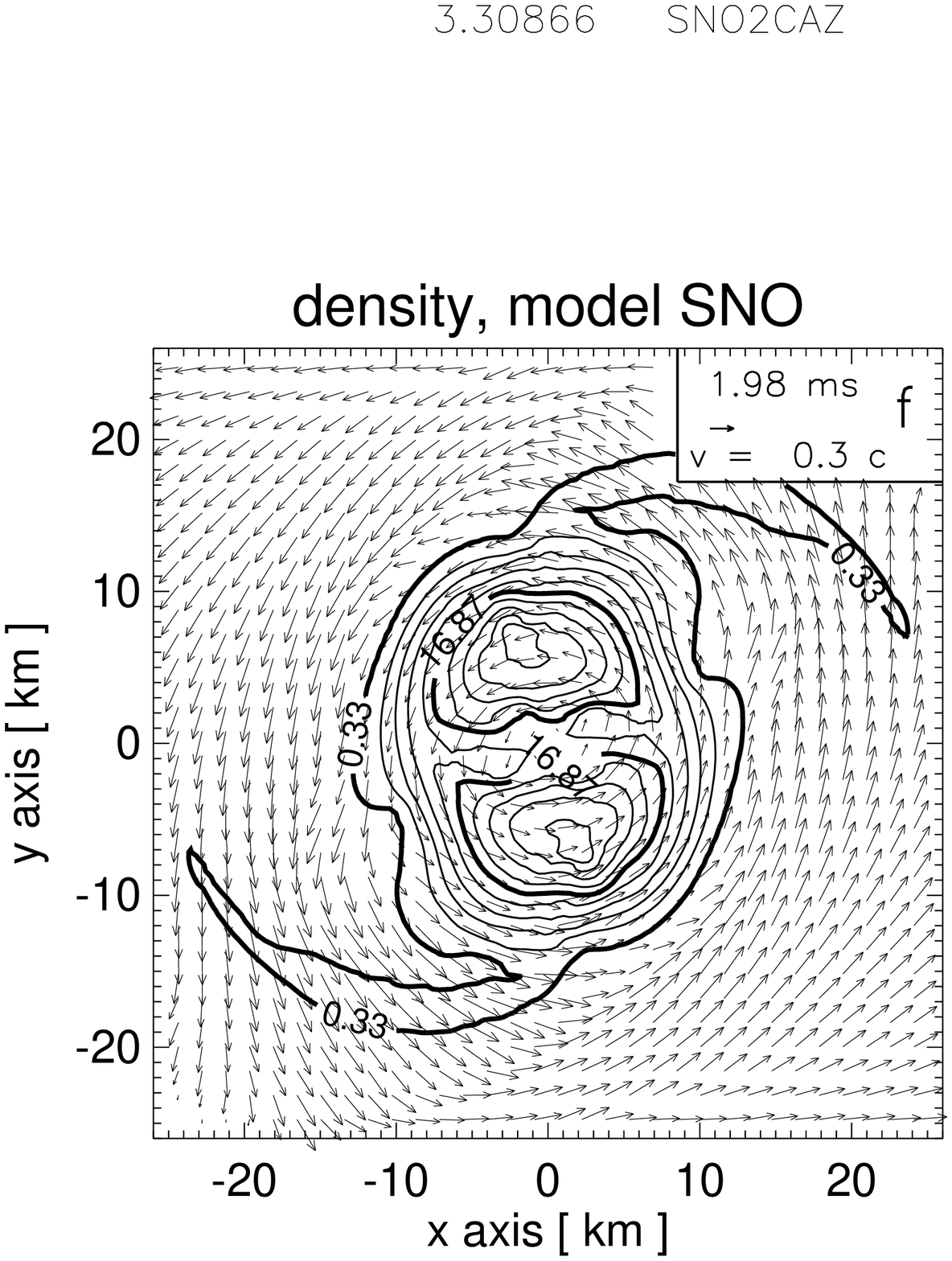} \\[-4ex]
 \end{tabular}
\caption[]{Cuts in the orbital plane of Model~SNO at six instants in
time showing the density contours together with the velocity field.
The density contours are linearly 
spaced in intervals of 0.1$\rho_{\rm max}$
starting at 0.01$\rho_{\rm max}$ and are labeled in units 
of $10^{14}$g/cm$^3$.
The legend at the top right corner of each panel gives the scale of
the velocity vectors and the time elapsed since the beginning of the
simulation
}
\label{fig:SNOcont}
\end{figure*}

We can only speculate why our Model~ZCM wobbles around and oscillates
for a longer time than the merger in the simulation of 
Zhuge et al.~(1994).
To calculate their model they used only 1024~SPH particles, which
is at least an order of magnitude less than the number of zones
resolving the central part of the merged object in our Model~ZCM.
Thus it seems possible that in Zhuge et al.~(1994) a larger numerical
viscosity of the SPH~code damps out the quadrupolar pulsations more
quickly than in our calculations.
Another possible cause for the different oscillation behaviour of our
Model~ZCM compared to Zhuge et al.~(1994) might be the differences 
of the phy\-sical description, namely the 
lack of tidal deformations in the initial state and the implementation
of backreaction terms in Model~ZCM.
The fact that Model~ZCM does not initially have any tidal bulges and
thus is not in equilibrium with the gravitational field will lead
directly to oscillations of the neutron stars reflected e.g.~in the
values of the maximum density.
This has been observed clearly in previous models, e.g.~see Fig.~10
in Ruffert et al.~(1996), in which one sees that the oscillations have
a period of one dynamical time (i.e.~the fundamental mode) and damp
out in roughly 10~periods.
Although the wobbling noticed in model~ZCM might conceivably be
triggered by this initial disequilibrium, the former is surely not
sustained by the latter: the wobbles continue for a time over which
the initial oscillations are damped out.
The effect of the different implementation of the gravitational wave
backreaction is more difficult to assess, but in general it has a
damping effect.
Initially, when the neutron stars are well separated, the backreaction
terms are responsible for the decay of the orbit, i.e.~the spiral-in
of the neutron stars.
This phase is simulated with similar effect by~Model~ZCM and 
Zhuge et al.~(1994), since without gravitational waves the neutron
stars would circle each other practically indefinitely.
However, during the final plunge and the subsequent evolution 
Zhuge et al.~(1994) do not include any backreaction.
Once the neutron stars have come close enough, hydrodynamic
effects alone dominate the merging process (cf.~Rasio \& Shapiro 1994
and references therein).
Our Model~ZCM is thus subjected to more damping during the final
stages of the merging process than in Zhuge et al.~(1994), so this
again cannot explain the longer-term wobbling.

\section{Results of Model SNO\label{sec:resultsSNO}}

\subsection{Initial conditions\label{sec:initSNO}}

In Model~SNO we use the same mass and radius for the neutron stars and
the same adiabatic exponent ($\Gamma=2$) as Shibata et al.~(1992)
for their Model~III.
Thus, we place the two neutron stars with a mass of $M=1.4~M_\odot$
each and a radius of 9~km at an initial distance of 27~km.
The two main differences between our run and the one of
Shibata et al.~(1992) are (1) the hydrodynamical integrator, 
and (2) the initial rotational state of the neutron stars and the
correspondingly constructed equilibrium configuration. 
Although the algorithm to calculate the gravitational wave
backreaction terms is the same in 
both works, we integrate the hydrodynamic equations with the PPM
scheme, in contrast to Shibata et al.~(1992) who employed a finite
difference method that is {\it not} based on a Riemann-solver.
In a number of previous models Nakamura \& Oohara~(1991) considered 
neutron stars at rest in a corotating system which is equal to a
solid-body type rotation in an inertial frame.
Contrary to that, Model~III of Shibata et al.~(1992) was constructed
as a ``spinning'' model, because 
the neutron stars were given spins in the corotating frame.
Corresponding to this neutron star rotation
Shibata et al.~(1992) assumed axisymmetric neutron stars in
rotational equilibrium initially, but did not solve for the equilibrium
configuration in the common gravitational potential, arguing 
that tidal forces are much smaller than self-gravity.
However, with the velocity distribution chosen by 
Shibata et al.~(1992) the initial velocities of all parts of the
neutron stars are collinear in an inertial frame, say parallel to the
$y$-axis, and the neutron stars do not rotate.
We therefore construct our neutron star models with the same velocity
distrubution as unperturbed spherical polytropes, whereby we also
assume that tidal forces are small compared to self-gravity.
We consider it to be more plausible that, if no
spin is present in the inertial system, spin rotational deformation
does not need to be taken into account either.

In their publication Shibata et al.~(1992) show the density
distribution and flow pattern as functions of time for Model~III as
well as the gravitational wave forms and the gravitational wave
luminosity. 
We shall therefore concentrate on these quantities in the comparison
of Model~III of Shibata et al.~(1992) and our Model~SNO.

\begin{figure}
 \epsfxsize=8.8cm \epsfclipon \epsffile{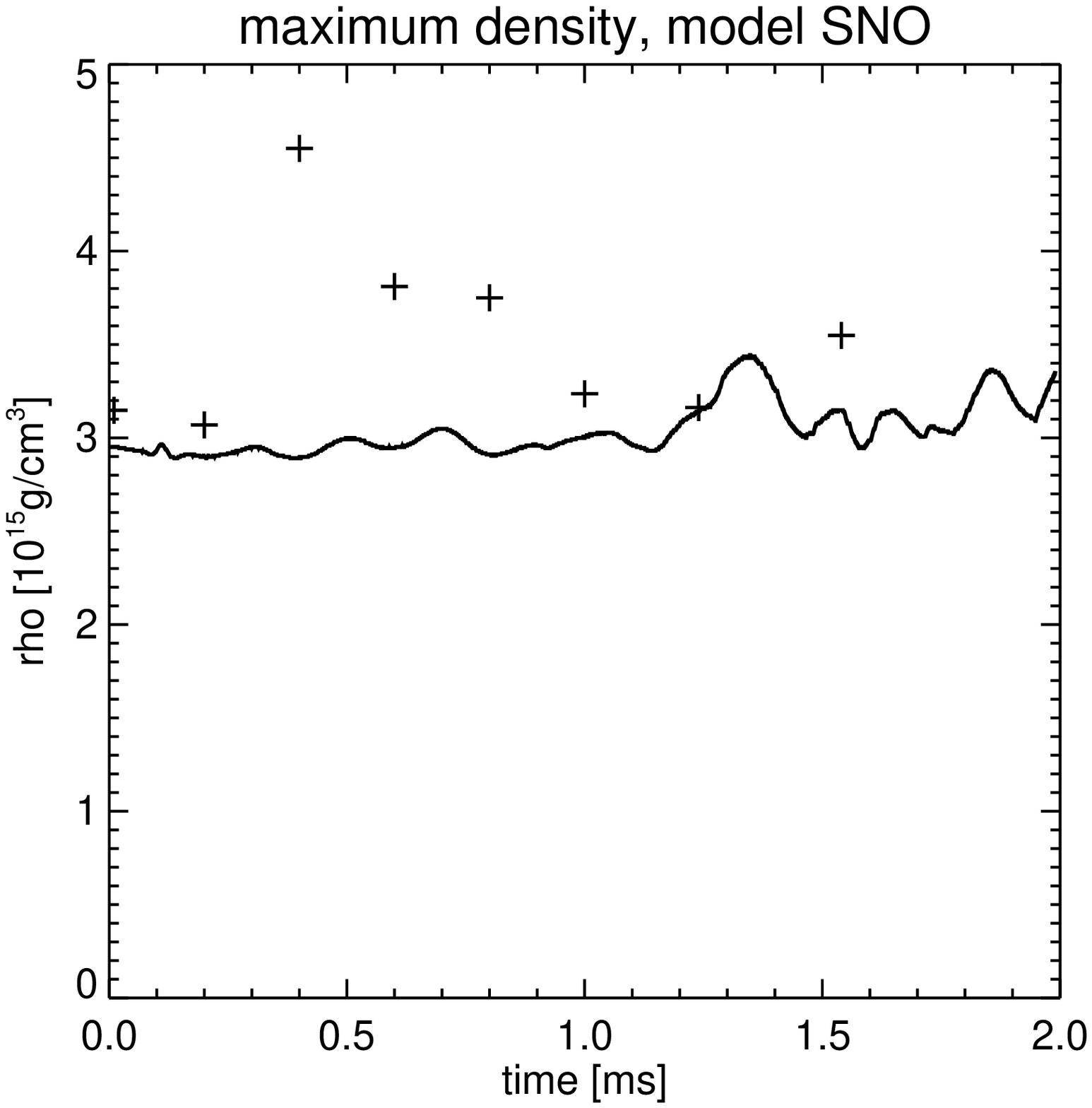}
\caption[]{The maximum density on the grid as a function of time for
Model~SNO (solid line).
The maximum density values as deduced from Fig.~3 of Shibata et al.~(1992)
are indicated by crosses
}
\label{fig:SNOmaxrho}
\end{figure}

\begin{figure}
 \epsfxsize=8.8cm \epsfclipon \epsffile{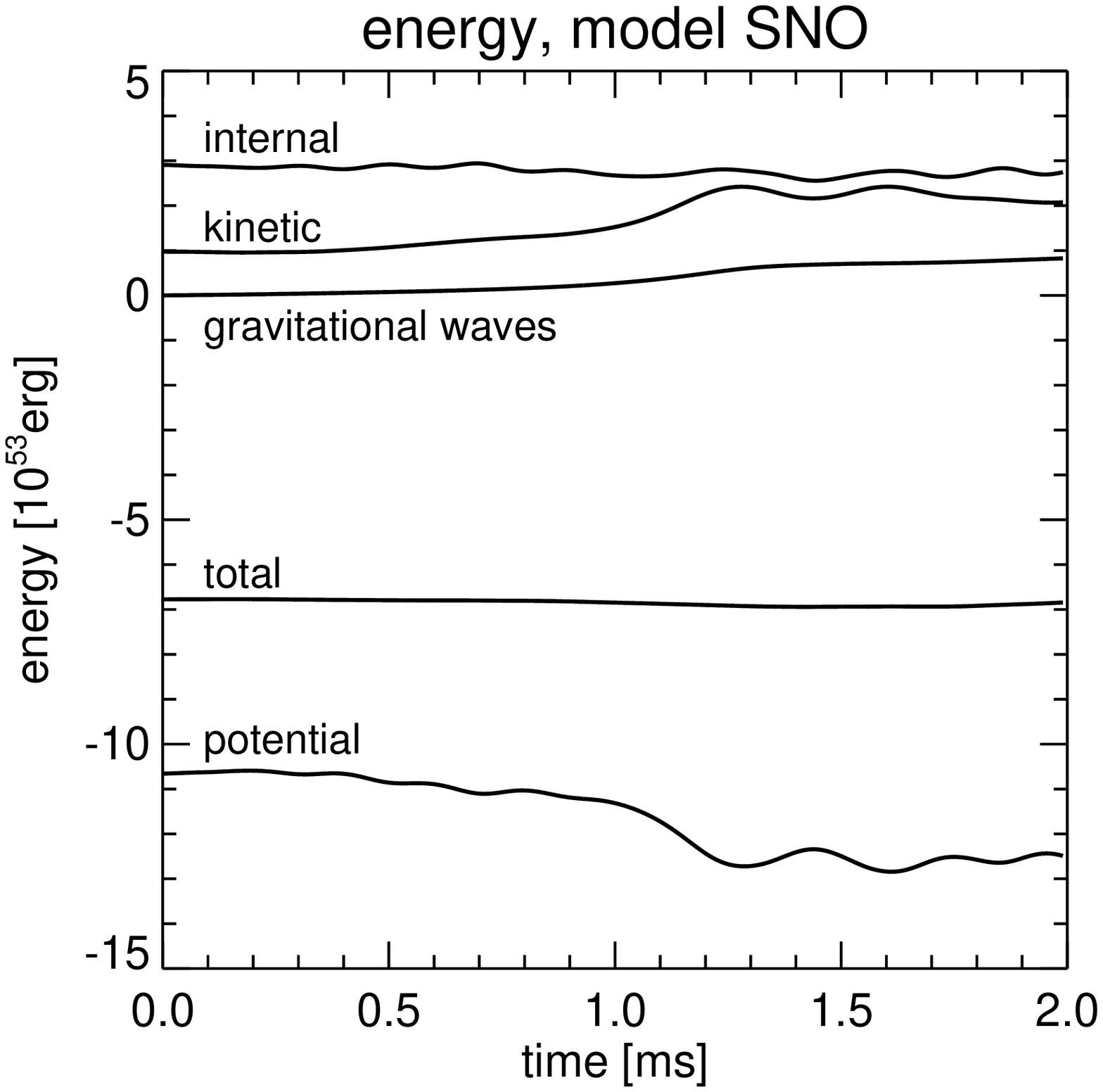}
\caption[]{Kinetic energy, internal energy, gravitational potential
energy, and emitted gravitational wave energy as functions of time for
Model~SNO. The total energy contains all individual energies
}
\label{fig:SNOenergy}
\end{figure}

\begin{figure}
 \epsfxsize=8.8cm \epsfclipon \epsffile{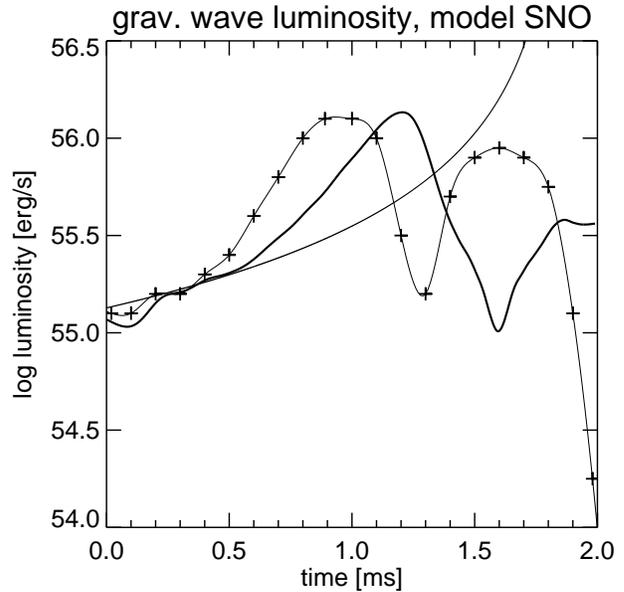}
\caption[]{The gravitational wave luminosity as a function of time for
Model~SNO (bold).
The crosses represent the values taken from Fig.~6b of 
Shibata et al.~(1992).
They are connected by splines.
The monotonically rising curve represents the luminosity of a
point-mass binary
}
\label{fig:SNOgrlum}
\end{figure}

\begin{figure}
\epsfxsize=8.8cm \epsfclipon \epsffile{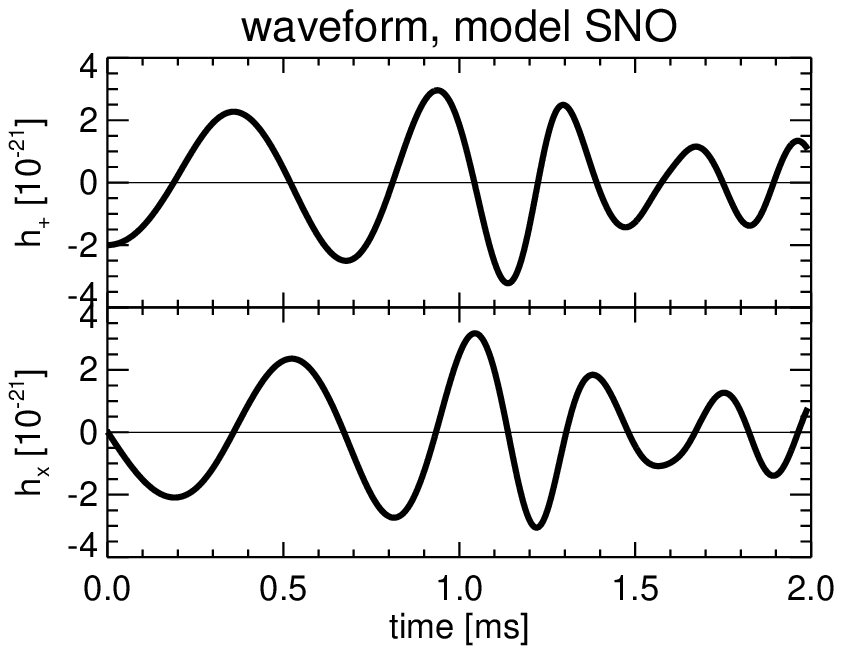}
\caption[]{The gravitational wave forms, $h_+$ and $h_\times$,
for Model~SNO as observed on the z-axis at 10~Mpc distance}
\label{fig:SNOform}
\end{figure}

\subsection{Dynamical evolution\label{sec:dynSNO}}

Snapshots of the density distribution in the orbital plane together
with the velocity field are shown at six instants in
Fig.~\ref{fig:SNOcont}.
The panels could, in principle, be compared with the 
panels of Fig.~3 of Shibata et al.~(1992), respectively.
In practice, however, the direct comparison of these contour plots is
difficult, unfortunately, since Shibata et al.~(1992) do not use
contour values that are constant in time and, moreover, they plot
arrows to indicate the flow field, but do not specify
the actual magnitude of the fluid velocities.
We used their prescription for the values of the density contours
in Fig.~\ref{fig:SNOcont}.

One can recognize that initially, at~$t\approx0.4$~ms and
at~$t\approx0.8$~ms, our Model~SNO is still in agreement with the one 
of Shibata et al.~(1992).
However, at~$t\approx1.2$~ms two distinct minima appear between the
cores of the neutron stars in Model~SNO, while in Model~III of 
Shibata et al.~(1992) the most prominent minimum is at the 
center in between the neutron stars.
These two minima of our Model~SNO merge to form the central minimum
at $t\approx1.6$~ms, while two secondary minima are additionally
present to the sides of the central minimum.
Thus Model~SNO at $t\approx1.6$~ms resembles Model~III, albeit at a
different time, $t\approx1.2$~ms.
This indicates a shift in time by roughly $\Delta t\approx0.3$~ms
of the temporal evolution of both
models that will become more clear in the next section.
The three minima between the neutron star cores merge again, forming
one elongated string by $t\approx1.8$~ms.
The structure of the merged object with two sub-cores remains present
until the end of our simulations at $t\approx2$~ms, while the 
density valley between becomes less prominent.

A quantity better suited for comparison is the maximum density as a
function of time, the values of which can be found inserted in the
panels of Fig.~3 of Shibata et al.~(1992).
In Fig.~\ref{fig:SNOmaxrho} these values are displayed at eight points
in time, together with the maximum density for our Model~SNO. 
The maximum density of Model~III by Shibata et al.~(1992) peaks
at around $t\approx0.4$~ms which is different from our Model~SNO. 
It is hard to explain this difference of the density evolution without
having more detailed information about Model~III of 
Shibata et al.~(1992).
At the time the density maximum occurs in the latter model, the
neutron stars are still clearly separated, in both
their Model~III (Fig.~3c in Shibata et al.~1992) and in our Model~SNO
(Fig.~\ref{fig:SNOmaxrho}).
There is an indication that at $t\approx0.4$~ms the density
distribution in the neutron stars of Shibata et al.~(1992) develops a
``crack'' or ``discontinuity''.
The ``crack'' seems to emanate from the center and propagate along the
direction of the $x$- and $y$-axes in both neutron stars.
We speculate that this might cause the higher density
maximum at $t=0.4$~ms in the model of Shibata et al.~(1992).
The reason for the differences in the evolution of the maximum density
and the neutron star separation seems not to be associated with the
initial shape and structure of the neutron stars due to the different
assumptions about the neutron star spins.
Rather than differences of the mass dis\-tri\-bu\-tions, the different
momenta of the stars might be responsible for the discrepant evolution.

In Fig.~\ref{fig:SNOenergy} we show the temporal evolution of some
integral energy quantities: the potential, kinetic, internal energies
as well as the energy in gravitational waves and the total sum of
all energies.
The total energy is conserved to 1\% compared to the maximum potential
energy.
The more compact structure of the merged object is reflected in the
deeper potential at the end of the simulation.
Note the slight oscillation present especially in the potential and
kinetic energies which shows the wobbling of the sub-cores.
The internal energy hardly changes indicating that the merging process
proceeds rather softly without strong shocks.

\begin{figure*}
 \begin{tabular}{cc}
  \epsfxsize=8.8cm \epsfclipon \epsffile{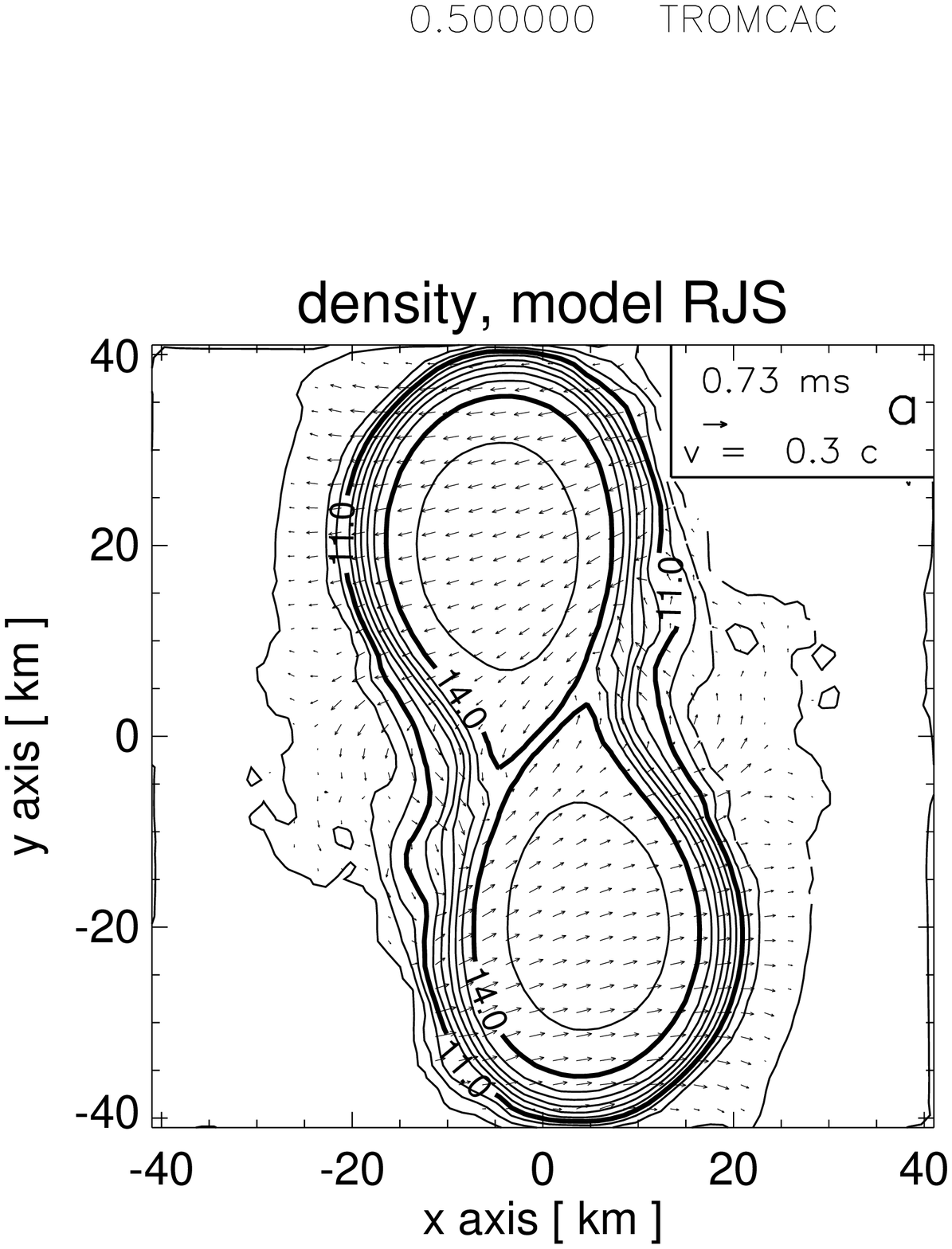} & 
  \epsfxsize=8.8cm \epsfclipon \epsffile{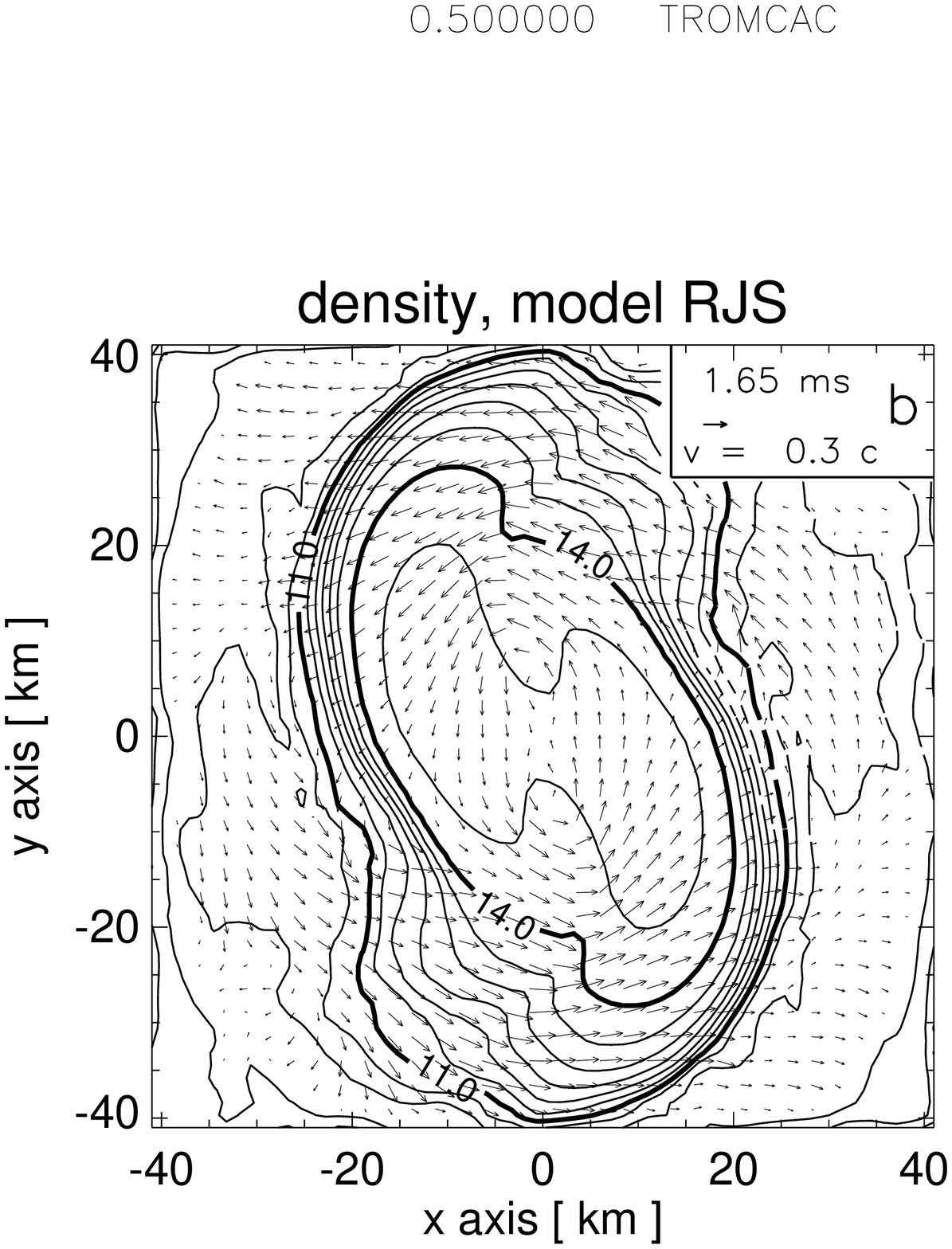} \\[-2ex]
  \epsfxsize=8.8cm \epsfclipon \epsffile{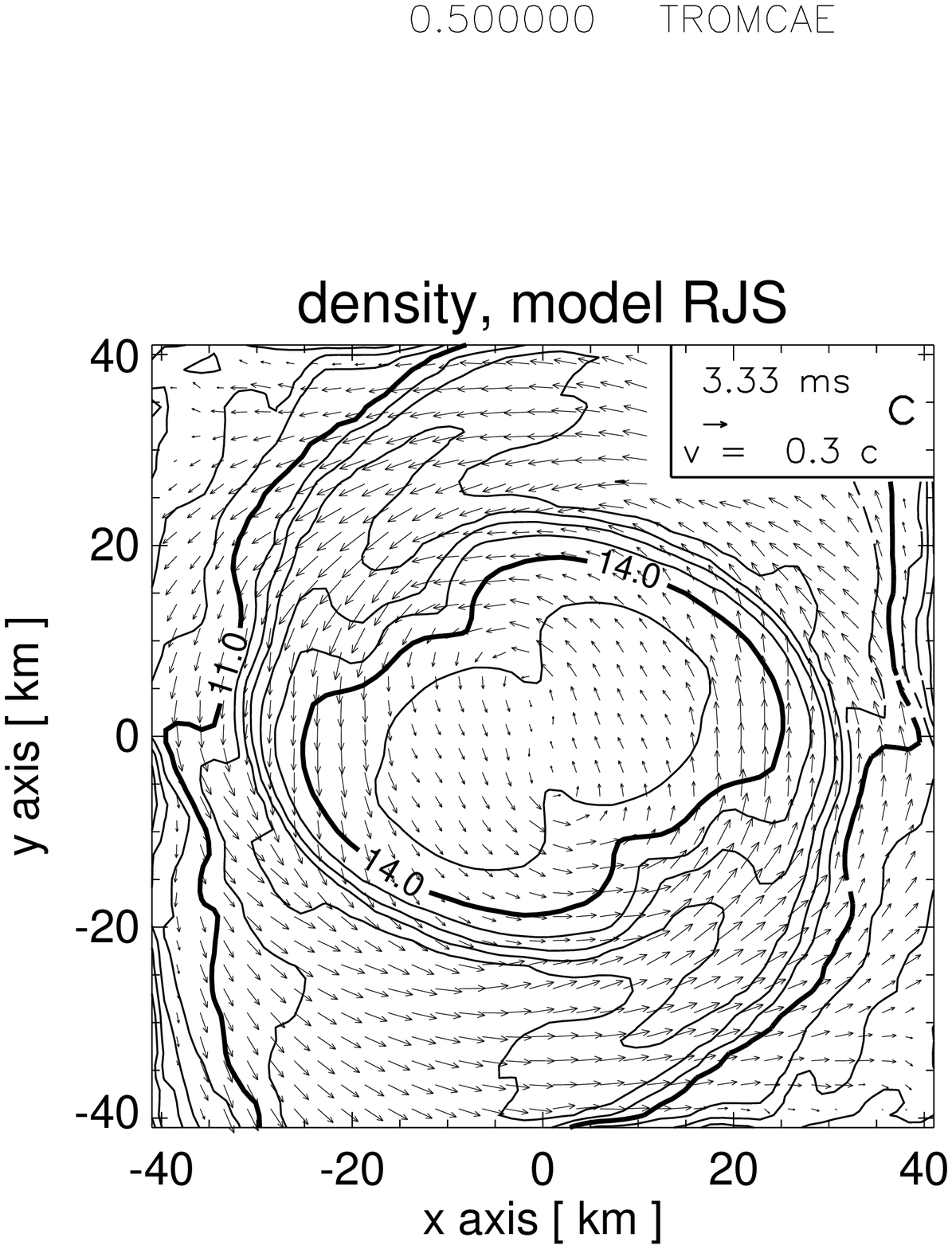} & 
  \epsfxsize=8.8cm \epsfclipon \epsffile{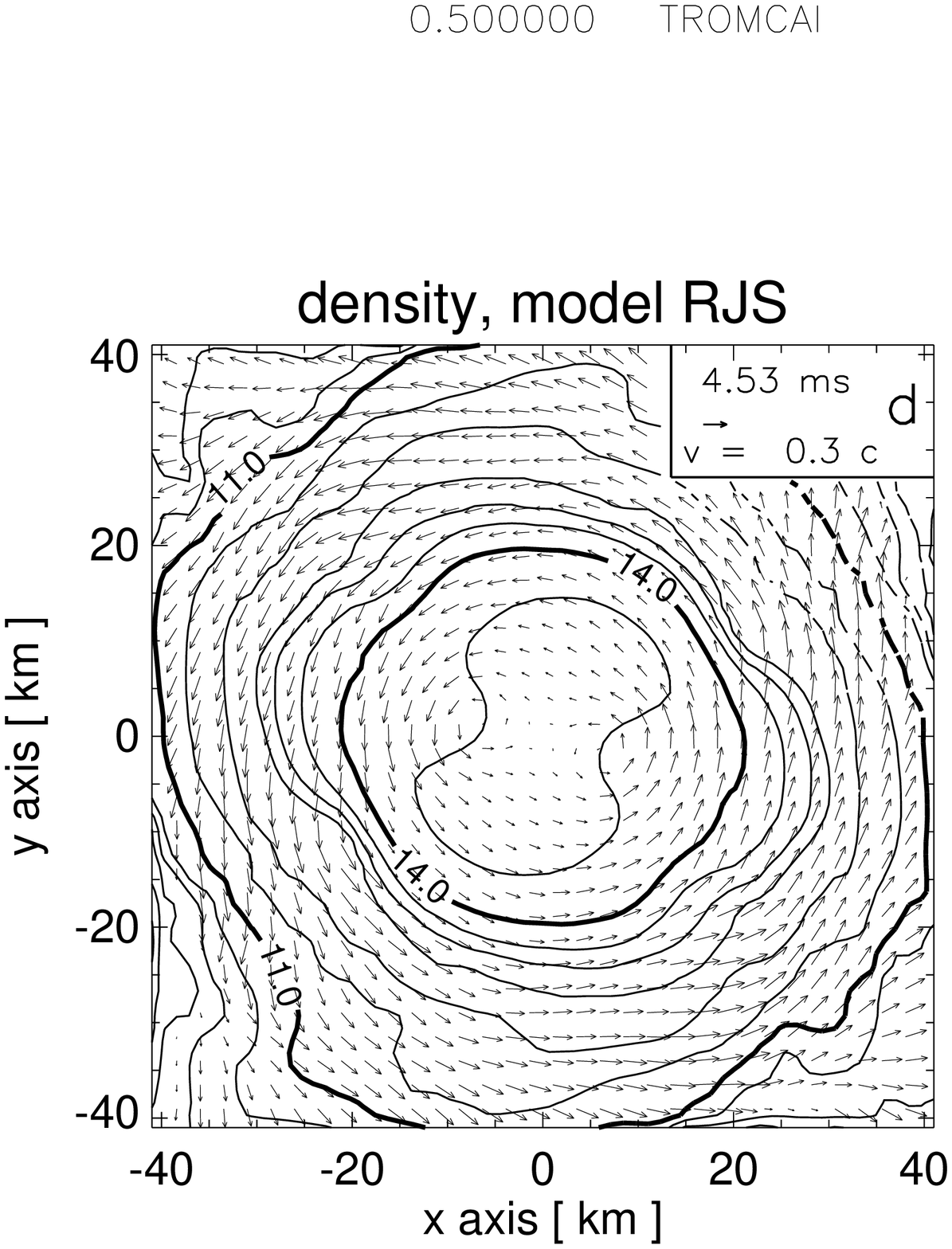} \\[-4ex]
 \end{tabular}
\caption[]{Cuts in the orbital plane of Model~RJS at four instants in
time showing the density contours together with the velocity field.
The density contours are logarithmically spaced with intervals
of 0.5~dex. The density is measured in units of g/cm$^3$.
The bold contours are labeled with their respective values.
The legend at the top right corner of each panel gives the scale of
the velocity vectors and the time elapsed since the beginning of the
simulation
}
\label{fig:RJScont}
\end{figure*}

\subsection{Gravitational wave forms and luminosity}

Figure~\ref{fig:SNOgrlum} displays the gravitational wave luminosities
as functions of time for Model~SNO and for Model~III of
Shibata et al.~(1992; Fig.~6b).
The maximum of the emission (which is located in the model of Shibata
et al.~(1992) at $t\approx0.9$~ms) is shifted in time 
by $\Delta t\approx 0.3$~ms, but agrees in height to within 10\%.
Note that model~SNO follows the point-mass binary approximation for
roughly 0.2~ms longer than Model~III.
We assume that (a) this initial deviation of Model~III is primarily
responsible for the subsequent time shift visible in the density
contours and the gravitational wave luminosity, and that (b) the
initial deviation is due to the differing initial rotational states of
the neutron stars.
Since the separation at the beginning of the simulations is very close
to the dynamic stability limit (cf.~Lai et al., 1994, and references
cited therein) a small difference in density distribution can cause a
substantial time lag: if the neutron stars are initially further 
separated than the
stability limit, gravitational wave emission has to continue for some
time to drive the decay of the orbit before the dynamical instability
can set in.

For completeness we also show in Fig.~\ref{fig:SNOform} the
gravitational wave forms for Model~SNO which can
be compared with Fig.~6a of Shibata et al.~(1992).
Due to the time shift mentioned above, the signal of our Model~SNO is
roughly one wave period shorter at the end of the simulation than
Model~III. 

\section{Results of Model RJS\label{sec:resultsRJS}}

\subsection{Initial conditions}

Model~RJS is calculated with neutron stars which have
the same mass, radius, and central density as those in Model~A64 of 
Ruffert et al.~(1996). 
Using the Lane-Emden equation and the mass--radius relation, this was
achieved by appropriately adjusting the polytropic constant~$K$ and
the adiabatic exponent $\Gamma$ (cf.~Table~\ref{tab:models}) for the
equation of state. 
Placing the neutron stars at the same initial distance of~42~km and
giving them the same velocity distribution as in as Model~A64, we are
left with a model differing only in the employed equation of state:
Model~RJS uses a polytopic equation of state according to
Eq.~\ref{eq:Eeos}, while Model~A64 was calculated with the 
equation of state of Lattimer \& Swesty~(1991) and included the
effects due to neutrino emission.

Out of the variety of quantities shown and discussed in 
Ruffert et al.~(1996), we decided to compare the evolution of the mass
distribution, the separation of the density maxima as a function of
time to show possible differences in the dynamics, and the
luminosity, wave form, and energy spectrum of the gravitational wave
emission.

\begin{figure}
 \epsfxsize=8.8cm \epsfclipon \epsffile{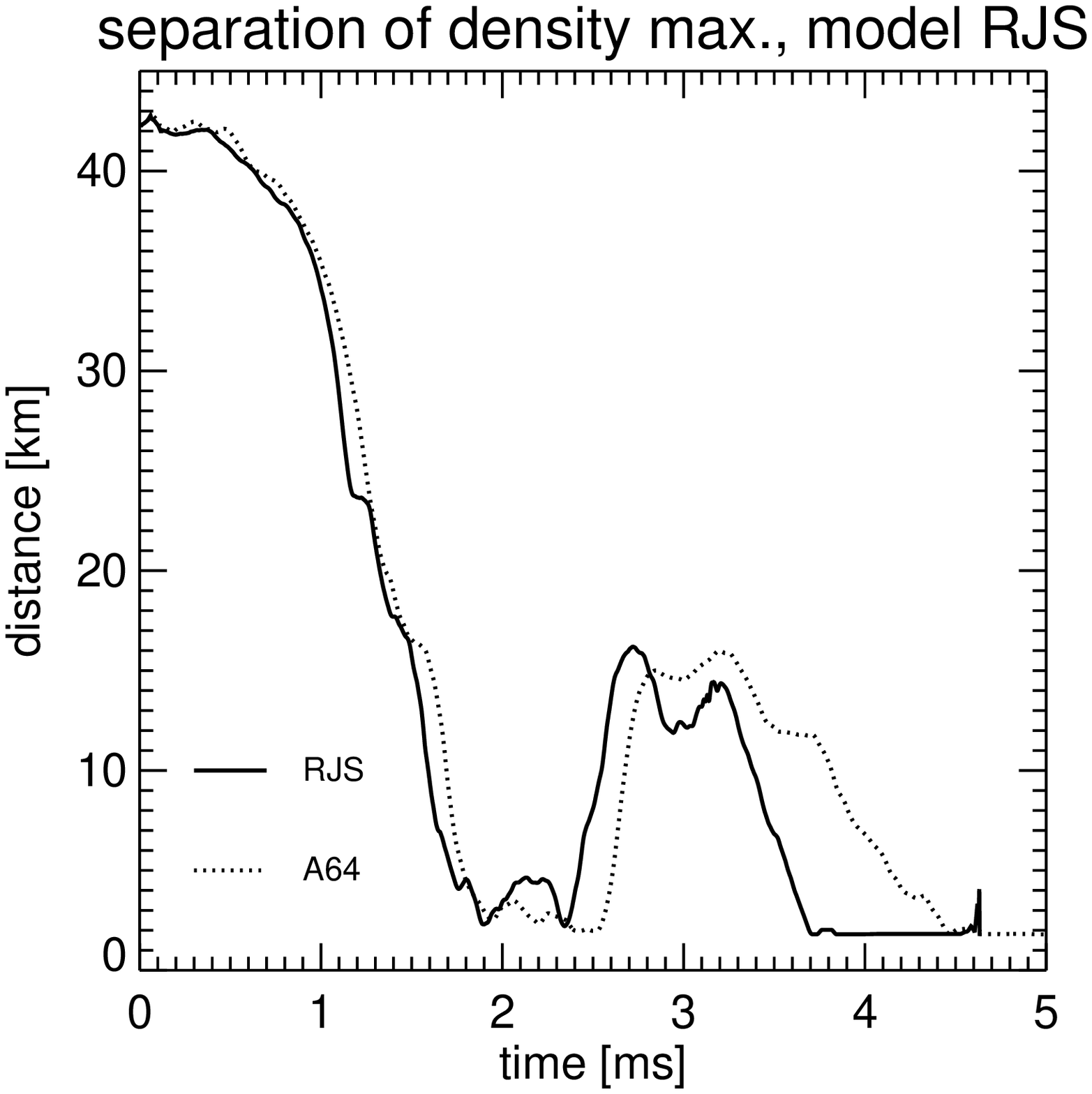}
\caption[]{The separation of the density maxima of the two neutron
stars as a function of time for Model~RJS (bold) and for Model~A64 (dotted)
of Ruffert et al.~(1996)
}
\label{fig:RJSsepden}
\end{figure}

\begin{figure}
 \epsfxsize=8.8cm \epsfclipon \epsffile{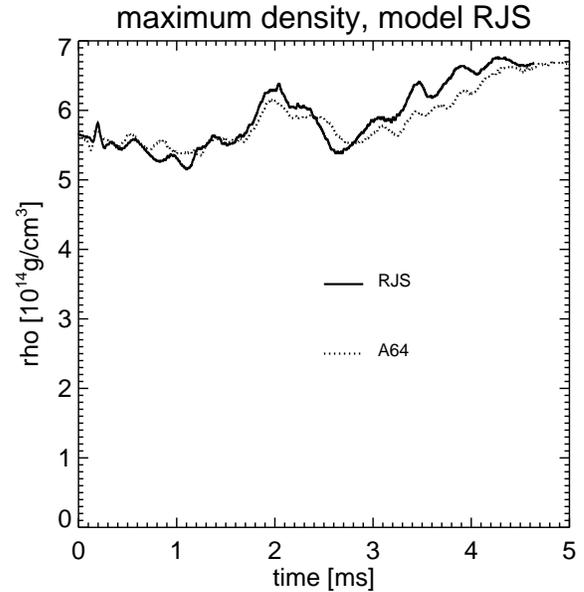}
\caption[]{The maximum density on the grid as a function of time for
Model~RJS (bold) and for Model~A64 (dotted) of Ruffert et al.~(1996)
}
\label{fig:RJSmaxrho}
\end{figure}

\begin{figure}
 \epsfxsize=8.8cm \epsfclipon \epsffile{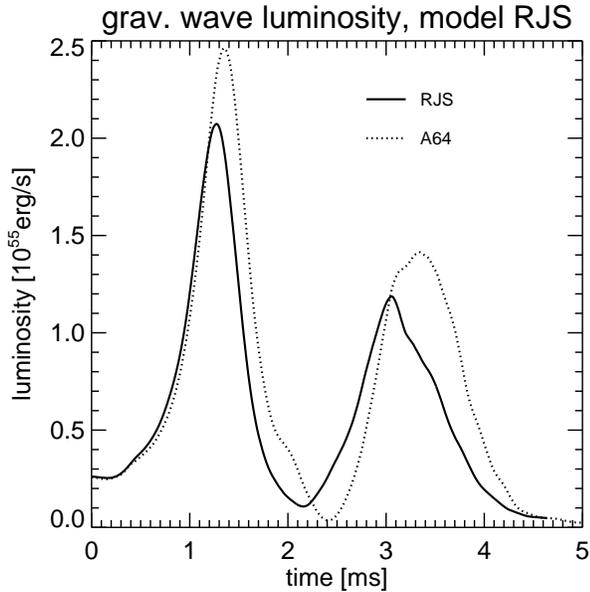}
\caption[]{The gravitational wave luminosity as a function of time for
Model~RJS (bold) and Model~A64 (dotted) of Ruffert et al.~(1996)
}
\label{fig:RJSgrlum}
\end{figure}

\begin{figure}
 \epsfxsize=8.8cm \epsfclipon \epsffile{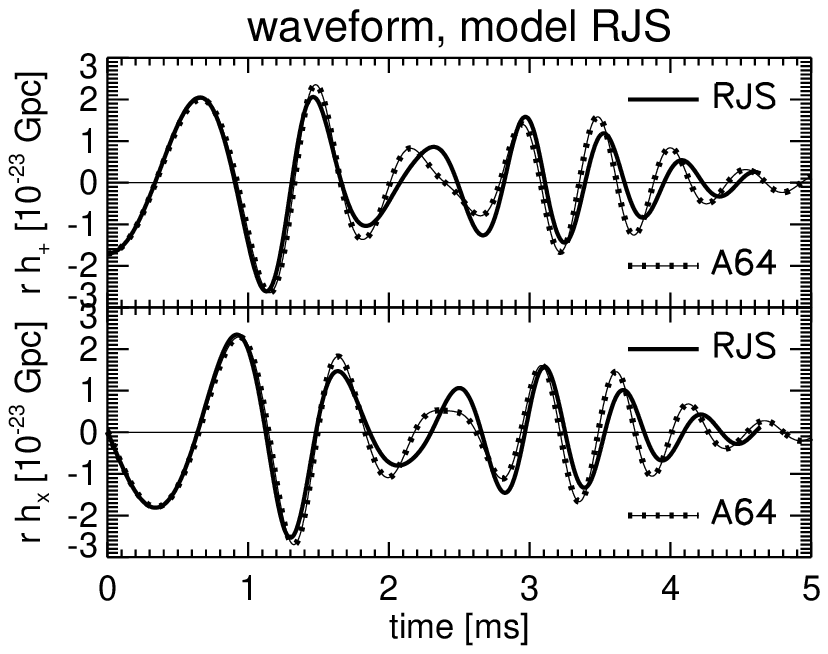}
\caption[]{The gravitational wave forms, $h_+$ and $h_\times$,
for Model~RJS (bold) and Model~A64 (dotted) of Ruffert et al.~(1996)
}
\label{fig:RJSform}
\end{figure}

\subsection{Dynamical evolution}

The contour plots in Fig.~\ref{fig:RJScont} show the dynamical
evolution.
The mass distribution is visualized by density contours with arrows 
superimposed to indicate the flow velocities.
These plots can directly be compared with the Figs.~4c, 4e, 5a, and~5c
of Ruffert et al.~(1996), respectively.
Initially, until about $t\approx1.7$~ms, only minor differences occur.
At $t\approx1.7$~ms the neutron star surfaces towards the downstream
sides (at the positions $x\approx0$~km, $y\approx\pm20$~km) are
more extended in Model~RJS (Fig.~\ref{fig:RJScont}b) than in
Model~A64 of Ruffert et al.~(1996) (see Fig.~4e there).
This effect also can be seen in Fig.~\ref{fig:RJScont}c where the
contour for the density value of $10^{11}$~g/cm$^3$ extends further
out than the marginal ``spiral arm'' visible in Fig.~5a in 
Ruffert et al.~(1996).

Compared to the models computed with the equation of state of Lattimer
\& Swesty (1991) in Ruffert et al.~(1996), all three models discussed
here, ZCM, SNO, and~RJS, have a less steep density decline at the
neutron star surface and develop a more extended disk after merging.
This general phenomenon is explained by the stiffness of the
polytropic equations of state which have adiabatic exponents between 2
and 2.32.
This is much stiffer than the equation of state of Lattimer \&
Swesty~(1991) in the subnuclear regime, where $\Gamma\approx1.33$ to 1.66.
An increase of the internal energy, e.g.~by the dissipation of kinetic
energy in shocks or via shear effects, which is important in the
regions close to the surface, therefore, leads to a strong pressure
increase and thus to an inflation of the stellar layers involved.

In the core region of the merged object, Figs.~\ref{fig:RJScont}c
and~\ref{fig:RJScont}d reveal more fine structure than Model~A64 of 
Ruffert et al.~(1996; Fig.~5a and~5c) which indicates
a stronger damping of high-frequency oscillation modes in the latter
model, caused by the bulk viscosity of the physical equation of state
used there.
At time $t\approx4.5$~ms both simulations show a rather similar
overall structure of the merger (compare Fig.~\ref{fig:RJScont}d to
Fig.~5c in Ruffert et al.~1996).
In both cases the innermost density contour 
($\rho\approx3\cdot10^{14}$~g/cm$^3$) is dumb-bell shaped and the
underlying double-core structure means that the merging process is
still going on.
Also during this phase the exterior layers at subnuclear densities are
significantly more extended in Model~RJS due to the stiffer polytropic
equation of state.

On the whole, the coalescence in Model~RJS proceeds very similarly to
Model~A64 of Ruffert et al.~(1996).
Figs.~\ref{fig:RJSsepden} and~\ref{fig:RJSmaxrho} show the separation
of the density maxima and the value of the maximum density as
functions of time for the two models.
In both models the neutron star density maxima merge within 2~ms and
then separate again for a short while when the double-core structure
forms (between 2.5~ms and about 3.5 to 4~ms) because of the large 
angular momentum of the merged object.
Model~RJS comes to ``rest'' faster (by roughly 0.5~ms) than Model~A64 
which has separated density maxima from 2.5~ms to 4.5~ms.
This difference might be caused by a larger angular momentum transport
outwards through the more extended disk of Model~RJS. 
This hypothesis is strongly supported by the outwardly directed
velocity field
which can be seen in Fig.~\ref{fig:RJScont}d, indicating mass
and angular momentum loss accross the grid boundaries.
Notice also that due to the differences of the subnuclear equation of
state the transition from the central core of the merged object to the
disk region is less sharp in Model~RJS (compare Fig.~\ref{fig:RJScont}b
with Fig.~5c in Ruffert et al.~1996).

Before the neutron stars merge, their central density 
decreases slightly by about 5\% (see Fig.~\ref{fig:RJSmaxrho}) and
during the coalescence the maximum density increases  by about 20\% of
the initial value.
The overall dynamical behaviour, reflected by the separation of the
density maxima and the value of the maximum density, 
reveals much less difference between the models than can be seen from
the density contours.
The similar dynamics
can easily be understood because the construction of the initial
neutron stars and the adjustment of the parameters~$K$ and~$\Gamma$ 
in the polytropic equation of state Eq.~\ref{eq:Peos} should ensure
very similar behaviour (e.g~response to compression)
of the polytropic equation of state and the
Lattimer \& Swesty~(1991) equation of state in the nuclear and
supranuclear regimes.
The differences occur in the outer regions of the neutron stars at
densities below the nuclear matter density where the polytropic
equation of state for the chosen value of $\Gamma$ is much stiffer
than the equation of state of Lattimer \& Swesty~(1991).
This difference mainly concerns the evolution of the outer layers and
thus the formation, structure and properties of the disk that forms
during merging. 
Since, however, only a comparatively small fraction of the neutron
star mass has densities less than nuclear matter density, the overall
dynamical evolution of the bulk of the matter is unaffected by the
difference of the low-density equation of state.
From the gradual increase of the density maximum we conclude that the
merging is as smooth as in Model~ZCM and not so violent as in Model~SNO
where the density increases rapidly during the final coalescence 
(see Fig.~\ref{fig:SNOmaxrho}).

\subsection{Gravitational wave forms and luminosity}

The gravitational wave luminosity is a very sensitive indicator of
differences in the overall dynamical evolution and 
the mass distributions, since it
involves the third time derivative of the quadrupole moment.
One recognizes from Fig.~\ref{fig:RJSgrlum} that the main 
maximum of the gravitational
wave luminosity is lower by 20\% and slightly narrower in
Model~RJS compared to Model~A64.
Roughly the same statements apply to the second maximum.
These differences between the two models are concordant with the
temporal behaviour of the separation of the density maxima in both
models (see Fig.~\ref{fig:RJSsepden}), but are much smaller
than the differences between Models~A64, B64, and~C64 of
Ruffert et al.~(1996).

The wave forms (Fig.~\ref{fig:RJSform}) show only minor differences:
both the amplitudes and the phases of the signals coincide well.
Slight discrepancies develop after $t\approx2$~ms which
corresponds to the time after the first merging of the density
maxima has occurred (Fig.~\ref{fig:RJSsepden}) and the gravitational
wave emission reaches a maximum (Fig.~\ref{fig:RJSgrlum}).
Once the separate density maxima have vanished, the merged
configuration has loses its strongly quadrupolar deformation and
therefore radiates gravitational waves less efficiently.

In agreement with the temporal displacement of the evolution visible
in Figs.~\ref{fig:RJSsepden}, \ref{fig:RJSmaxrho}, 
and~\ref{fig:RJSgrlum}, a small phase shift of the gravitational wave
forms after $t\approx2$~ms is caused by the slightly faster rotation
of Model~A64.
The higher angular momentum of the merged configuration (see above)
together with the more compact structure of the layers at
intermediate densities (between $10^{12}$~g/cm$^3$ and
$10^{14}$~g/cm$^3$) due to the less stiff equation of state, is also
the reason for somewhat larger gravitational wave amplitudes and the
enhanced gravitational wave luminosity in case of Model~A64 at times
later than about $t\approx3$~ms. 
The total energy emitted in gravitational waves in Model~RJS is
$2.8\cdot10^{52}$~erg, roughly 20\% less than the
$3.4\cdot10^{52}$~erg emitted by Model~A64 within 5~ms (taken from
Fig.~23 of Ruffert et al.~1996).

\subsection{Gravitational wave spectrum}

\begin{figure*}
 \begin{tabular}{cc}
 \put(2.5,1.6){{\large $t=1.60$ms}}
  \epsfxsize=8.8cm \epsfclipon \epsffile{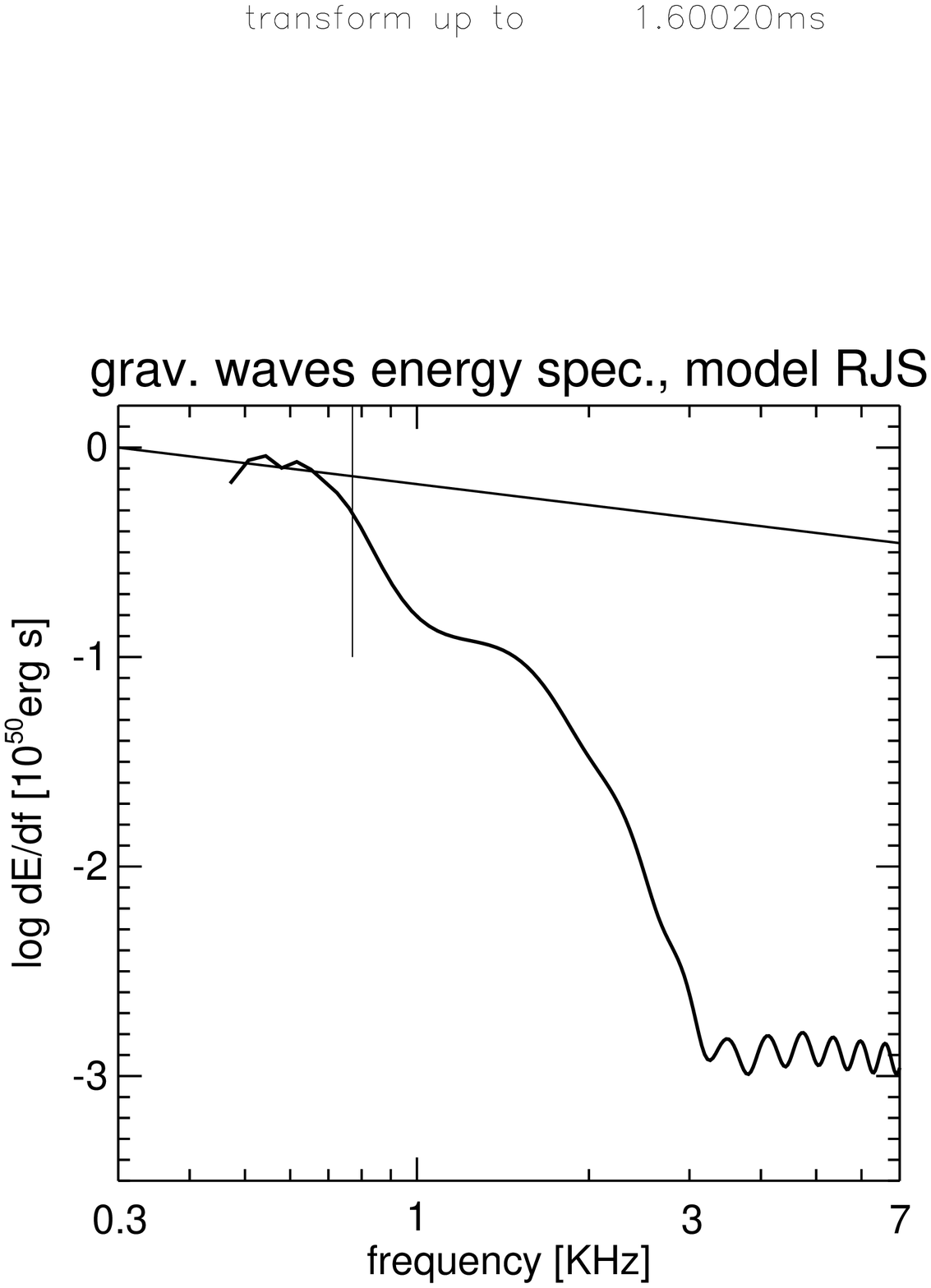} & 
 \put(2.5,1.6){{\large $t=3.20$ms}}
  \epsfxsize=8.8cm \epsfclipon \epsffile{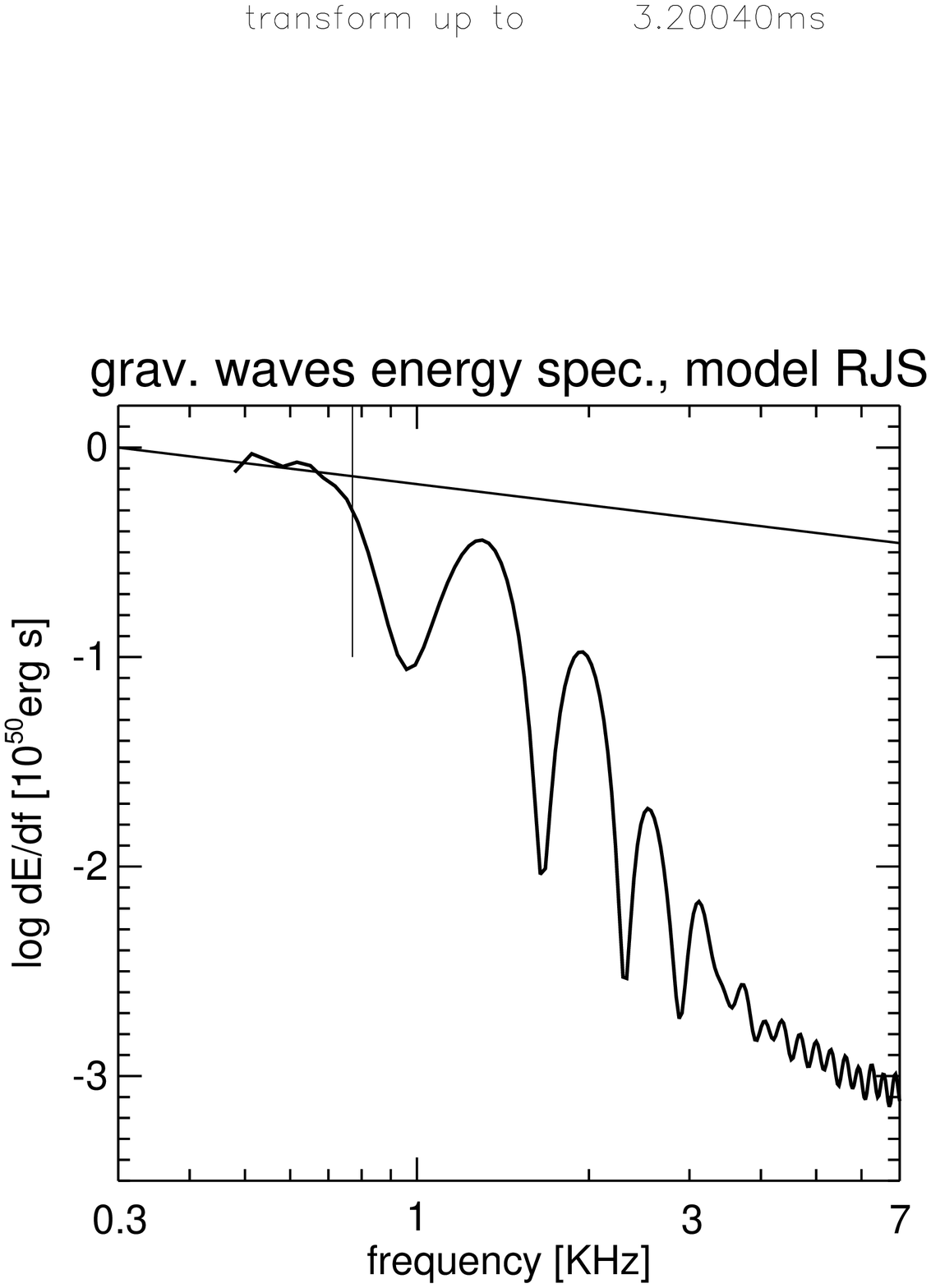} 
      \\[-2ex]
 \put(2.5,1.6){{\large $t=4.64$ms}}
  \epsfxsize=8.8cm \epsfclipon \epsffile{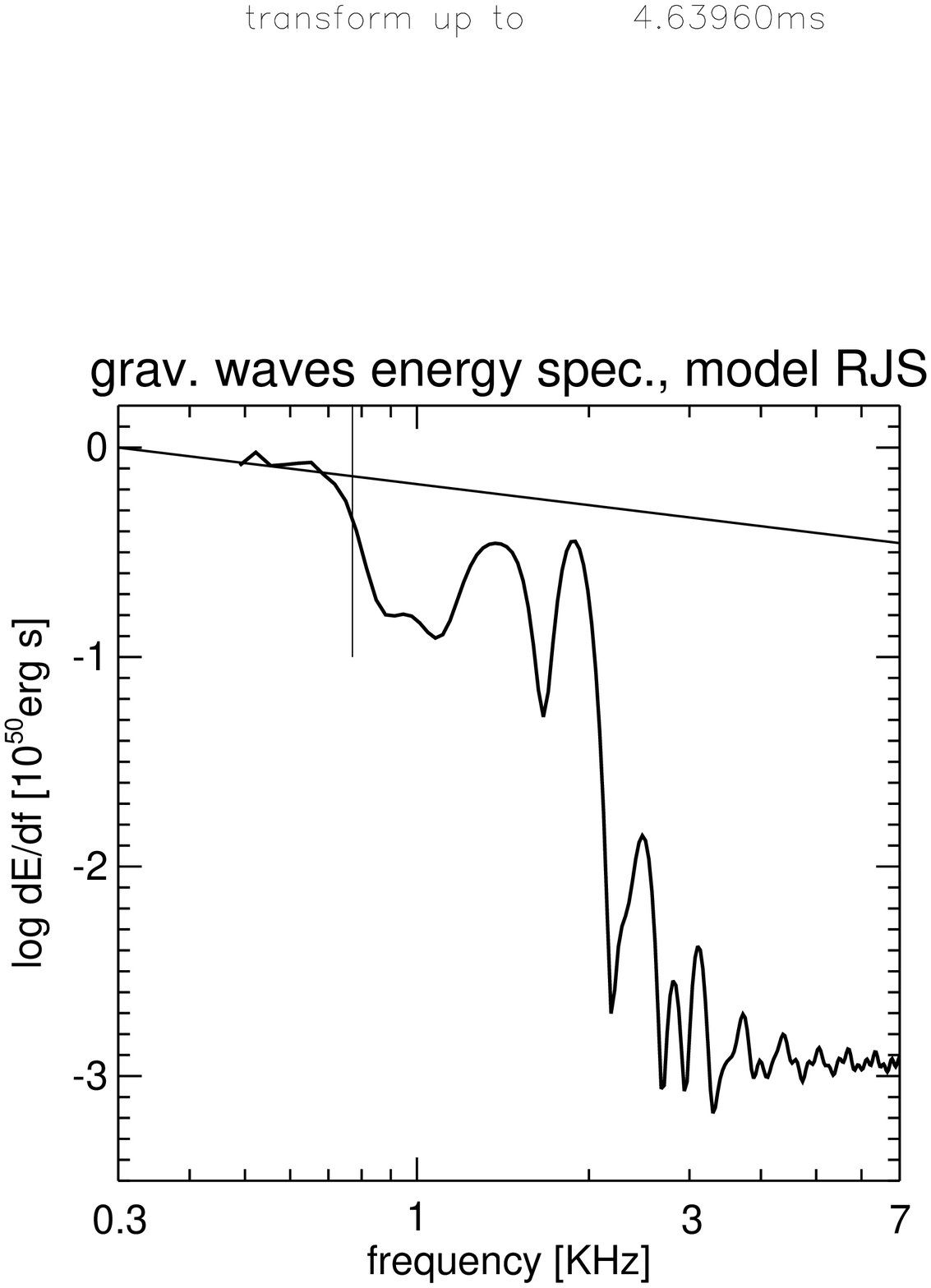} & 
\raisebox{8cm}{\parbox[t]{8.8cm}{
\caption[]{Energy spectrum of gravi\-ta\-tional waves emitted in
Model~RJS. 
The times until which the Fourier transforms are performed are
indicated in  the lower left corners of the panels.
They correspond to the snapshots of the density distributions shown in
Fig.~28 of Ruffert et al.~(1996).
The straight, downward sloping line is the spectrum of a point-mass
binary.
The vertical lines indicate the frequency corresponding to the orbital
frequency of two point masses at the initial distance of the neutron
stars in our numerical model
}\label{fig:RJSspec}
}}
 \end{tabular}
\end{figure*}

The close resemblance of the gravi\-ta\-tional wave forms
(Fig.~\ref{fig:RJSform}) is mirrored in the gravitational wave spectra
which share the main features in the two models~RJS and~A64 from
Ruffert et al.~(1996).
The spectra of Model~RJS can be found in Fig.~\ref{fig:RJSspec} and
should be compared with Fig.~28 in Ruffert et al.~(1996).
They contain the same kind of information as already described in the
context of Fig.~\ref{fig:ZCMspec}, i.e.~the 
frequency distribution of the gravitational wave energy emitted until
the times shown in the lower left corners of the plots, together with 
a line representing the energy loss per unit frequency interval of a
point-mass binary.

At time $t\approx3.2$~ms the spectra look very similar and
Models~RJS and~A64 both have a spectral minimum at $f\approx1$~KHz 
and spectral maxima at $f\approx1.3$~KHz and $f\approx2.0$~KHz.
Even for frequencies $f>2.0$~KHz the decline of the spectra is
pretty much the same.
The major visible differences are the strengths of the second and of
the third spectral maximum at $f\approx2.4$~KHz.
While the maximum at $f\approx2$~KHz is more pronounced in Model~RJS,
the peak at $f\approx2.4$~KHz is higher than in Model~A64. 
In both models the $f\approx1.3$~KHz maximum is mainly generated before
$t\approx3.2$~ms, while the $f\approx2$~KHz maximum grows strongly
after $t\approx3.2$~ms.
The $f\approx2$~KHz maximum tends to shift to $f\approx1.8$~KHz 
(more clearly in Model~RJS) between
$t\approx3.2$~ms and $t\approx4.64$~ms and becomes slightly more 
prominent in Model~RJS than in Model~A64.
After $t\approx4.6$~ms the spectral peaks at frequencies between
about~2~KHz and~2.4~KHz have merged into one structure in Model~A64,
while in Model~RJS there are still two sharp, well separated maxima
and the $f\approx2.4$~KHz feature has not gained much strength any
more.
We think that these results can again be explained by the stronger
pulsational and vibrational activity of Model~A64 after merging which
adds to the power of the gravitational wave emission at frequencies
around and above $f\ga2$~KHz.

\section{Discussion and summary\label{sec:end}}

The hydrodynamic simulations presented in this work follow the
dynamical evolution and the gravitational wave emission of two merging
neutron stars modeled as polytropes.
We compared our results with three models published earlier and tried
to obtain information about uncertainties and differences caused by
the numerical scheme employed for integrating the hydrodynamic
equations, by the numerical resolution and setup of the initial model,
and by the use of a polytropic equation of state instead of the
physical equation of state of Lattimer \& Swesty~(1991).
We chose the following three models for comparison:
(a) ``Run~2'' of Zhuge et al.~(1994), 
(b) ``Model~III'' of Shibata et al.~(1992), and
(c) ``Model~A64'' of Ruffert et al.~(1996).
We attempted to construct the initial configurations as similar to
these models as possible and also used polytropic equations of state
which were the same or which guaranteed to reproduce the properties
(mass, central density, radius) of the neutron stars.

\subsection{Model ZCM}

Run~2 of Zhuge et al.~(1994) is different from our Model~ZCM in (a)
one numerical aspect: Zhuge et al.~(1994) used an SPH
algorithm while we employed the grid-based PPM scheme and (b) two
physical aspects: (i) the gravitational wave backreaction is
prescribed in an {\it ad hoc} way by Zhuge et al.~(1994), because it
is included in the computation when the neutron stars are still
separate but switched off during the merging process; (ii) both
computations start with different
initial separations and tidal deformations.
Our Model~ZCM seems to retain a ``double core'' structure and thus a
large quadrupolar deformation during the
merging for a longer time than Run~2.
This has two consequences: On the one hand, 
the secondary and tertiary maxima of the gravitational wave luminosity
are significantly higher in our calculation, and on the other hand, 
the peak in the gravitational wave spectrum at a frequency of 
$f\approx1.8$~KHz is more prominent, too
(a first maximum is present at about 1.3~KHz).
Both effects are caused by a more extended post-coalescence period in
Model~ZCM where the merger performs ``ringing'' and ``wobbling''
motions and emits more power in gravitational waves at frequencies
around $f\approx1.8$~KHz.
We find a local minimum of the gravitational wave spectrum at a
frequency of about $f\approx1.5$~KHz in good agreement with Run~2 of
Zhuge et al.~(1994), although this minimum is not quite as deep in our
calculation. 
The integrated gravitational wave energy in both models is also nearly
the same, implying that the stronger emission of our Model~ZCM at high
frequencies is accompanied by a deficit at lower frequencies, which we
indeed find in the band $f\approx0.7$--1.1~KHz.
We suppose that these differences are mainly caused by the larger
numerical viscosity of the SPH~code and the comparatively poor
resolution with only $1024$~particles in the computation of 
Zhuge et al.~(1994).
Although the two main physical differences (initial tidal
deformations, backreaction implementation) could in principle be also 
partly responsible for the wobbling, we assume their effect to be
small: (i) the initial disequilibrium due to tidal deformations has been
shown (Ruffert et al.~1996) to be damped out within a time shorter
than the ringing, and (ii) the gravitational backreaction has a
damping effect in the longer-term merging process.

\subsection{Model SNO}

Shibata et al.~(1992) performed their calculations with a finite
difference grid-based explicit Eulerian code just as we do, albeit
with a different hydrodynamical integrator, a second order advection
using a method proposed by LeBlanc (see Appendix~A of Oohara \&
Nakamura~1989), which does not make use of a Riemann solver.
They incorporated the gravitational wave backreaction in a way
conceptually equivalent to our treatment.
Their grid had $121^3$~zones.
To solve the Poisson equation, the Incomplete Cholesky decomposition
and Conjugate Gradient (ICCG) method by Meijerink \& van der Vorst
(1977) was used.
Contrary to this we implemented a fast Fourier method and our grids
had a size of 128$\times$128 (and 32 zones vertical to the orbital
plane). 
However, long before any hydrodynamical interaction between the
neutron stars occur, at $t\approx0.4$~ms when the neutron stars
just barely touch each other, the maximum density on the grid showed
an extreme value in the calculations of Shibata et al.~(1992) which we
were unable to reproduce in our Model~SNO.
This difference is difficult to explain and
we can only speculate that it might be associated with the choice of
Shibata et al.~(1992) to use configurations in rotational equilibrium
for the two neutron stars which were given spins in the frame that
corotates with their orbital revolution. 
Because these spins
of the stars were absent for an observer in an inertial
frame and because Shibata et al.~(1992) did not construct equilibrium
configurations in the common gravitational field of the two stars
(which would make more sense), we did not follow their choice but
started our simulations as usual with spherical neutron star models.
This obvious discrepancy causes an initial delay of the merging
by $\Delta t\approx0.3$~ms in our model~SNO compared to Model~III
of Shibata et al.~(1992).
Nevertheless, the gravitational wave luminosity amplitude of their
model and of our simulation coincide surprisingly well to within 10\%.

\subsection{Model RJS}

In a third comparative study we performed a simulation, Model~RJS,
with initial conditions (neutron star masses, radii, central
densities, initial separation, and velocities) like those used in
Model~A64 of Ruffert et al.~(1996).
The only major difference was that instead of the physcial equation of
state of Lattimer \& Swesty (1991) --- which also allowed us to
include neutrino physics in the computation --- we used a
polytropic equation of state in Model~RJS with the initial value of
the parameter $K$ and the constant adiabatic index $\Gamma$ adjusted
such that the initial neutron star properties are reproduced.

We found the overall dynamical evolution during the coalescence,
e.g.~timescale of merging, formation of spiral-arm-like structures of
matter spun off the neutron star surfaces by tidal and centrifugal
forces, post-merging oscillations and pulsations, to be very similar
in both models.
This is particularly clearly seen in the behaviour of the neutron star
separation and of the maximum density on the grid as functions of
time.
Correspondingly, the gravitational wave luminosity, gravitational wave
form, and gravitational wave spectrum share the main structural
features.
The peak gravitational wave luminosity and the total energy emitted in 
gravitational waves are only about 20\% smaller in the polytropic
model.
The gravitational wave forms are very similar but have the tendency to
develop a slight phase shift after the two neutron stars have merged
into one object.
These minor differences reflect the fact that the properties
of the Lattimer \& Swesty~(1991) equation of state in the nuclear and
supranuclear regimes --- which are most important for the neutron
star structure as well as for the gravitational wave signature during
the merger --- are sufficiently well reproduced.
This is ensured by our determination of the equation of state
parameters~$K$ and~$\Gamma$ by prescribing the same mass, central
density and radii of the neutron star models constructed with the
polytropic equation of state and with the Lattimer \& Swesty~(1991)
equation of state.

The minor differences of both calculations, the roughly 20\% smaller
peak emission of gravitational waves in the polytropic case and the
small phase shift of the gravitational wave forms, are mainly caused
by differences of the equations of state in the subnuclear regime
($\rho\la10^{14}$~g/cm$^3$).
In this regime
in particular, the description of the properties of neutron star
matter with an equation of state with a uniformly chosen adiabatic
exponent~$\Gamma$ is inadequate.
With the employed value of $\Gamma=2.319$ the equation of state
at subnuclear densities is by far too stiff.
As a result, the neutron star surface layers in Model~RJS were more
extended and the density decline less sharp compared to Model~A64 in
Ruffert et al.~(1996). 
Morerover, during merging the neutron star surfaces came into contact
earlier in Model~RJS which moderated the dynamical interaction of the
stars and reduced the luminosity outburst of gravitational waves
during this phase.
After merging, the overestimated stiffness of the equation of state led
to a slightly less compact central body of the merger in Model~RJS and
to a significantly more extended disk structure surrounding the
central, massive body than in Model~A64.
In particular, the properties of this disk, its mass, radial extent,
and rotational state,  therefore require the use of a proper physical
equation of state during the simulations of neutron star merging.

\begin{acknowledgements}
We acknowledge the indication of the possible relevance of the
presented comparison by an anonymous referee of the paper of 
Ruffert et al.~(1996).
The calculations were performed at the Rechenzentrum Garching on a
Cray~J90~8/512.
\end{acknowledgements}

\end{document}